\documentstyle[prl,aps,eqsecnum,preprint,tighten,epsf,floats]{revtex}
\begin{document}
\title{Is the phase transition in the Heisenberg model described by the
$(2+\epsilon)$-expansion of the Nonlinear 
$\sigma$-Model?}
\author{Guillermo E. Castilla}
\address{Department of Physics, Brookhaven National Laboratory, Upton, NY 11973}
\author{Sudip Chakravarty}
\address{Department of Physics and Astronomy, University of California Los 
Angeles
\\ Los Angeles, CA 90095-1547}
\date{\today}
\maketitle
\begin{abstract}
Nonlinear $\sigma$-model is an ubiquitous model. In this paper, the $O(N)$ model where the
$N$-component spin is a  unit  vector, ${\bf S}^2=1$, is considered. The
stability of this model with respect to gradient operators $(\partial_{\mu}{\bf
S}\cdot
\partial_{\nu}{\bf S})^s$, where the degree  $s$ is arbitrary, is
discussed. Explicit two-loop calculations within the scheme
of $\epsilon$-expansion, where
$\epsilon=(d-2)$, leads to the surprising result  that these operators are
relevant. In fact, the relevancy increases with the degree $s$. We
argue that this phenomenon in the
$O(N)$-model actually reflects the failure of the perturbative analysis, that is,
the $(2+\epsilon)$-expansion. It is likely that
it is necessary to take into account non-perturbative effects if one wants to
describe the  phase transition of the Heisenberg model within the context of the
non-linear $\sigma$-model. Thus, uncritical use of the $(2+\epsilon)$-expansion
may be  misleading, especially for those cases for which there are
not many independent checks.
\end{abstract}
\newpage

\section{Introduction}

The nonlinear $\sigma$-model is an ubiquitous model that can 
describe systems ranging from quantum spins\cite{CHN} to disordered electronic
systems\cite{Local}. In the present paper we shall discuss this model with
$O(N)$ symmetry. A particularly attractive method to solve this model is the
$(2+\epsilon)$-expansion\cite{Heisenberg}, where the spatial dimensionality, $d$,
defines
$\epsilon=d-2$.  In this method, one takes advantage of the proximity of the
non-trivial fixed point to the  zero temperature fixed point in the limit
$\epsilon \to 0$. One then expands around the zero temperature fixed point  to
obtain information about the non-trivial fixed point. Such arguments are clearly
powerful, especially because there are not many explicit analytical techniques to
solve this model. The purpose of the present paper is to examine this method more
critically. 

We begin by showing that, in the most natural definition of the
problem, one must examine the role of gradient operators of degree higher than
two, although, by power-counting, the higher order gradient terms are irrelevant.
The task of this paper is to show that  when  fluctuations are taken into
account, the higher order gradient operators become relevant, more so as  the
number of gradients increases. The simplest interpretation is that the fixed
point is infinitely unstable, and the method fails. This surprising result
was first discovered in a related
model in the context of Anderson localization of an electron in a random
potential\cite{Kravstov}. However, we believe that this phenomenon in the
$O(N)$-model actually reflects the failure of the perturbative analysis, that is
of the $(2+\epsilon)$-expansion. To appreciate this conclusion, consider the
$O(3)$ Heisenberg model for which much is known 
from high temperature series expansion, accurate
Monte-Carlo calculations, and from the
$(4-d)$-expansion of the ${\bf \phi}^4$ field theory. None of these methods
give any hint of any pathological behavior.
Indeed,  Kehrein, Wegner and Pismak have shown that in the $N$-component
${\bf \phi}^{4}$ model the one-loop
contributions always make the canonically irrelevant operators
even more irrelevant\cite{Wegner0}.
Thus, it is very likely that
it is necessary to take into account non-perturbative effects if one wants to
describe this phase transition within the context of the non-linear
$\sigma$-model.  At the very least, the results obtained from this expansion
should be taken with some caution, especially in those cases in which there
are not many independent checks. 

To one-loop order, the
anomalous dimensions of the high gradient operators of the $O(N)$ model were
first calculated by Wegner\cite{Wegner1}. In an earlier paper\cite{thepaper}, we
briefly reported the corresponding two-loop results, and we showed that the
relevance of the high-gradient operators persists to two-loop order. Since the
method and the unexpected results may not be familiar, we have decided to give a
full account in the present paper. We also analyze the renormalization group
equations, and we discuss the physical consequences of the renormalization
group flows. 
 
The plan of the paper is as follows. In  Sec. II, we define the high gradient
operators, and, in Sec. III,  we set up the background field method. In Sec. IV,
we calculate the one-loop correction to the anomalous dimension
to illustrate the efficacy of the background field method. In Sec. V, we
present the two-loop calculation. In Sec. VI, we 
discuss the flows of the renormalization group equations.  Sec. VII contains
our conclusions, and there are four Appendices.

\section{High gradient operators}
It is useful to begin with a soft-spin model because, in many instances, the
non-linear $\sigma$-model arises from a microscopic situation in which both the
direction and the magnitude of the order parameter field is allowed to fluctuate. 
One
then argues that as long as the $O(N)$ symmetry is spontaneously broken, the
interactions between the goldstone modes are precisely those given by the
non-linear $\sigma$-model\cite{Weinberg0}. To be specific let us consider the
$O(N)$-invariant $\phi^4$ field theory  and explore how the high gradient
operators arise. For this field theory, the action is given by 
\begin{equation}
S[{\bf \phi}]=\int d^dx\left[{1\over 2}
(\partial_{\mu}{\bf \phi})^2+{1\over 2}
r{\bf \phi}^2+u({\bf \phi}^2)^2\right].
\end{equation}
Below the mean field transition, set 
\begin{equation}
\phi (x) =\rho(x) {\bf S}(x), \ \ {\bf S}^2(x)=1,
\end{equation}
and separate out the direction, ${\bf S}(x)$, of the order parameter field 
${\phi}(x)$ from its magnitude $\rho(x)$. If we write
\begin{equation}
\rho(x) = m + \Delta \rho(x) ,
\end{equation}
where $m$ is the mean field value of the magnitude of the
order parameter, not containing any loop corrections, {\it i.e.},
 fluctuations. Then
\begin{equation}
S[{\bf \phi}]=\int
 d^dx\left\{{1\over 2}(m+\Delta \rho(x))^2\left(\partial_{\mu}{\bf
S}\right)^2+{1\over 2}\left(\partial_{\mu}{
\Delta\rho}\right)^2+{1\over 2} r(m + \Delta \rho(x))^2+u(m + \Delta
\rho(x))^4\right\}
\end{equation}
In principle, we can integrate out the fluctuations of the magnitude of the
order parameter, $\Delta \rho(x)$.  Because $\rho(x)$ is a massive field, 
the effective action is local on scales larger than the inverse mass. It
is now clear that, for scales larger than the inverse mass or the 
correlation length, 
the effective action
will involve gradient operators to infinite order in a functional Taylor 
expansion.
From  symmetry, and  from the locality of the action on  scales larger
than the correlation length associated with the fluctuations of the magnitude of
the order parameter field, we find that the  effective action can be cast in the form
\begin{equation}
S_{\rm eff}=
\frac{1}{2T} \int d^d x \left\{ 
  (\partial_{\mu} \vec{\bf S})^2 +
 U_{4} (\partial_{\mu} \vec{\bf S})^4 +
 V_{4}(\partial_{\mu} \vec{\bf S}\cdot \partial_{\nu} \vec{\bf S})^2
+ \cdots \ 
    \right \}  \label{effective}
\end{equation}
where $T, U_4, V_4,\dots,$ are the coupling constants; we are measuring the 
temperature $T$ in units of the bare spin stiffness constant. 

From power-counting, the gradient terms with powers larger than 2 are 
irrelevant and are usually dropped. If $U_{2s}$,
for $s > 1$, denotes the  coupling constant associated with the operator 
$(\partial_{\mu}{\bf S})^{2s}$, the full scaling dimension, $y_{2s}$, of
$(U_{2s}/T)$ is defined by
\begin{equation}
y_{2s}=2(1-s)+\epsilon+{\rm loop\ corrections},
\end{equation} 
where $\epsilon = (d-2)$; the dimension 
$(1/T)$ is $\epsilon$, plus loop corrections. Hence, if we ignore the loop
corrections, the contribution of the operator $(\partial_{\mu}{\bf S})^{2s}$
vanishes as
$\Lambda^{-2s}$ as the momentum cutoff in the theory
$\Lambda\to\infty$, for small $\epsilon$. In contrast, the second gradient
term is relevant for $\epsilon > 0$ and marginal for $\epsilon=0$. 

The
neglect of the higher order derivative terms leads to the usual partition
function of the  the nonlinear sigma model:
\begin{equation}
Z=\int D{\bf S}\  \delta({\bf S}^{2}(x)-1) \exp \left(
   -\frac{1}{2T} \int d^{d}x (\partial_{\mu} {\bf S})^{2} \right) \label{Partition}.
\end{equation}
This is also the na\"{i}ve continuum limit of the lattice Heisenberg model for
which the Hamiltonian is 
\begin{equation}
H/T=-{1\over T}\sum_{<ij>} {\bf S}_i\cdot {\bf S}_j,
\end{equation}
where the temperature is measured in units of the bare spin stiffness constant.
There are well-known arguments that the lattice Heisenberg model is in the same 
universality class as the $N$-component $\phi^4$ field theory. Therefore, all three
models, the nonliner $\sigma$-model, the $N$-component $\phi^4$ field theory, and the
lattice Heisenberg model of $N$-component unit vector spins should belong to the same
universality class, sharing the same long-distance and critical properties. 

However, power counting does not necessarily determine the effect of 
the operators. 
One must examine how loop corrections, or fluctuations, affect the
picture.  For the non-linear $\sigma$ model,
it may appear unlikely that the
corrections to the dimensions of the high
derivative terms coming from fluctuations  can overcome 
the canonical dimension.  
However, we shall show that, for large $s$, not only 
the one-loop
correction to the canonical dimension
is sufficiently large to
render  the full dimension positive, but the two loop
correction is  even larger. 
So, to two-loop order,
the operators  $(\partial_{\mu}{\bf S})^{2s}, s > 1, $ are relevant,
contrary to the common description of the non-linear $\sigma$-model.

\section{The Riemannian manifold and the Background Field Method}

The present calculation is nearly
impossible without an efficient method. In this section, we discuss a
formalism that is efficient.  The first step is to express the action as an
invariant in the space of cosets ${O(N)/O(N-1)}$\cite{Friedan,Zinn}. The length on
this manifold is
$ds^2=d\sigma^2+(d\pi^i)^2$, but because ${\bf S}^2 =\sigma^2+\pi^2=1$, we can eliminate
$\sigma$ using
$\sigma d\sigma + \pi\cdot d\pi=0$.
Therefore, the line element is
\begin{equation}
ds^2=g_{ij}(\pi)d\pi^id\pi^j,
\end{equation}
where the metric, $g_{ij}(\pi)$, is
\begin{equation}
g_{ij}(\pi)=\delta_{ij}+\frac{\pi_i \pi_{j}}{1-{\pi}^2}
\label{Metric}.
\end{equation}
There is no unique way of choosing coordinates on this manifold. 
The set $\{\pi^i\}$ and the
set $\{\tilde\pi^i\}$ will produce two different metrics, but they are related by the
transformation equation
\begin{equation}
g_{ij}(\pi)={\partial\tilde\pi^k\over \partial \pi^i}
{\partial\tilde\pi^l\over \partial \pi^j}
g_{kl}(\tilde\pi).
\end{equation}
The action in Eq. (\ref{Partition}) can now be written as
\begin{equation}
S(\pi)=\frac{1}{2T}\int d^dx g_{ij}(\pi)\partial_{\mu}{\pi^{i}}\partial_{\mu}{\pi^{j}}.
\label{action}
\end{equation}
Similarly, the high derivative operators can be written as
\begin{equation}
(\partial_{\mu_{1}}{\bf S} \cdot \partial_{\nu_{1}} {\bf S}) \ldots
(\partial_{\mu_{s}}{\bf S} \cdot \partial_{\nu_{s}} {\bf S})
=
(g_{ij}\partial_{\mu_{1}}\pi^{i}\partial_{\nu_{1}}\pi^{j}) \ldots
(g_{pq}\partial_{\mu_{s}}\pi^{p}\partial_{\nu_{s}}\pi^{q})
\label{Newoper}
\end{equation}
Clearly,  the high derivative
 operators and the action 
in Eq. (\ref{Partition}) are invariant
under the reparametrizations of the sphere, since  
$\partial_{\mu}\pi^{i}$ transforms as a vector under reparametrization. The
invariant measure is
\begin{equation}
[D\pi]=\prod_i\prod_x\sqrt{g(\pi)}d\pi^i(x),
\end{equation}
where $g(\pi)={\rm det} (g_{ij})$.

We  briefly recall the background field method\cite{background}. 
The strength of this method
is that covariant expressions can be handled easily and the explicit covariance can be
maintained at each step of the calculation. 
This turns to be important for an efficient organization of the calculations to be
described. There is another  important 
reason for using this
method.  When the fluctuations around the background field are integrated out, 
operators arise that
are not invariant under the reparametrizations of the sphere. One obtains
operators that are proportional to the classical equation of motion
\begin{equation}
\frac{\delta S}{\delta \pi_{i}}  =   \partial_{\mu} \partial^{\mu} \pi^{i}+
\Gamma^{i}_{jk} \partial_{\mu}\pi^{j}\partial^{\mu}\pi^{k}; \label{eqofmot}
\end{equation}
where
\begin{equation}
\Gamma^{i}_{jk}= \frac{1}{2} g^{it}[
\partial_j g_{tk}+ \partial_k g_{tj}- \partial_t g_{jk}]
\label{Connect}
\end{equation}
are the Christoffel symbols. Since $\Gamma^{i}_{jk}$ does not transform as a 
tensor,
the equation
of motion is not invariant, and the operators proportional
to the equation of motion are  not invariant quantities.
These are the redundant operators\cite{Wegner2} that do not affect the critical 
properties  and can
be removed by the reparametrizations of the sphere. Let us consider the shift 
\begin{equation}
\pi \to \pi + \theta f(\pi),
\end{equation}
in the partition function. Here $f$ is a smooth function, and $\theta$ is an 
infinitesimal parameter. 
Then, the redundant operator, $O_{\rm red}$, is
\begin{equation}
O_{\rm red}=\int d^dx\left\{f{\delta S\over \delta \pi}-{\delta f\over \delta \pi}\right\}.
\end{equation}
The first term comes from the action and the second from the measure. 
As stated above, the redundant operators can be removed by a reparametrization.
We adopt
dimensional regularization and the minimal subtraction scheme
for the calculation of the renormalization constants.
In the dimensional regularization scheme, the contribution of 
the measure term can be set to zero, and the redundant operators disappear 
if the equation of motion is
satisfied, which is an advantage  of using the background field method.

The anomalous dimension of a composite operator $O(x)$ is computed by computing 
the divergences created by the insertion in the correlation function, defined by
\begin{equation}
\Gamma_O^{(n)} = \langle O(x)\pi(x_1)\cdots\pi(x_n)\rangle.
\end{equation}
The divergence of the correlation function is not only proportional to 
the inserted opeartor, but it is usually a linear combination of a number of other
operators that are said to mix with the inserted opeartor. The renormalized operator
$O_i^R$ is defined as
\begin{equation}
O_i^R= \sum_j Z_{ij}O_j,
\end{equation}
where $Z_{ij}$ are the renormalization constants. The renormalized correlation function 
$\Gamma^{(n)}_{O,R}$ in the momentum space is given by
\begin{equation}
\Gamma^{(n)}_{O_i,R}(q,p_i,g,\mu)=Z^{(\frac{n}{2})}
Z_{ij}\Gamma^{(n)}_{O_j}(q,p_i,g_0),
\end{equation}
where $g_0$ is the bare coupling, the temperature $T$, and $Z$ is the wave 
function renormalization of the field. This leads to the renormalization group equation
\begin{equation}
\left[\left(\mu{\partial \over \partial \mu}+\beta(g){\partial\over 
\partial g}-{n\over 2}\gamma(g)\right)\delta_{ij}+\gamma_{ij}(g)\right]
\Gamma^{(n)}_{O_j,R}(q,p_i,g,\mu)=0.
\end{equation}
The quantity $\gamma(g)$ is the anomalous dimension of the field, and 
$\gamma_{ij}^O$ 
is the anomalous dimension (matrix) corresponding to the operator $O$, and
it is defined by
\begin{equation}
\gamma_{ij}^O=-\sum_k (Z^{-1})_{ik}\left(\mu\frac{d}{d\mu}\right)Z_{kj} , 
\label{anomal}
\end{equation}
where the derivative is computed with fixed bare couplings. 
One then diagonalizes the matrix $\gamma^O$ at the fixed point $g=g^*$ and
obtains
\begin{equation}
\left[\mu{\partial \over \partial \mu}-{n\over 2}\eta+\gamma_{\alpha}(g^*)\right]
\Gamma^{(n)}_{O_{\alpha,R}}(q,p_i,g^*,\mu)=0,
\end{equation}
where $O_{\alpha}$ is an eigenoperator.
The quantities $\gamma_{\alpha}$ are the eigenvalues
 associated with the eigenoperators
  $O_{\alpha}$, and $\eta=\gamma(g^*)$ is the anomalous dimension of the
field $\pi(x)$. The
solution of this equation leads to the behavior
\begin{equation}
\lim_{\rho\to \infty}\Gamma^{(n)}_{O_{\alpha,R}}(q,p_i,g^*,\mu)=\rho^D\rho^{-y_{\alpha}}
\Gamma^{(n)}_{O_{\alpha,R}}(q/\rho,p_i/\rho),
\end{equation}
where $D$ is the usual dimension of the correlation function without 
the operator insertion, and $y_{\alpha}$
is given by
\begin{equation}
y_{\alpha}=d-[O_{\alpha}]-\gamma_{\alpha}(g^*)\label{yalpha},
\end{equation}
where $[O_{\alpha}]$ is the engineering dimension of the operator $O_{\alpha}$.
Of course, the largest eigenvalue of the matrix
 $\gamma_{ij}(g^*)$ controls the 
critical behavior of the operators $O_i$.

In the
background field method\cite{Alvar81}, the initial step is
to write the field
$\pi(x)$ in terms of  new fields: 
\begin{equation}
\pi(x)=\psi(x)+\eta(x).  \label{spl2}
\end{equation}
The field $\psi(x)$ is chosen to satisfy the
classical equation of motion,
and $\eta (x)$ represents  quantum fluctuations. If we substitute this
decomposition in the action and in the definition of the composite operator and
expand in powers of
$\eta$, the expansion is not manifestly covariant. To obtain a covariant expansion, 
the field  $\eta$ is written in terms of normal coordinates
$\xi^{i}$ defined by the connection $\Gamma_{ij}^{k}$ in Eq. (\ref{Connect}).
The normal coordinate $\xi^{i}$ is the tangent vector to the
geodesic of the field manifold between $\psi(x)$ and $\pi(x)$ with
 a magnitude equal to the geodesic 
distance.
In terms of the coordinates $\psi$ and $\pi$, the normal coordinates,
$\xi$, have the expansion\cite{Alvar81}
\begin{equation}
\xi=\pi-\psi+\frac{1}{2}\Gamma_{mn}^{i}(\pi-\psi)^{m} (\pi-\psi)^n
+\cdots
\end{equation}
A systematic method to find the normal coordinate expansion
of an arbitray operator
is given in Ref.\cite{Alvar81}.
First, we note that for the manifold $O(N)/O(N-1)$, $\Gamma_{ij}^k=\pi^kg_{ij}$,
and the curvature tensor
$R^m_{npq}$ is defined by the equation:
\begin{equation}
R^m_{npq}=\partial_p\Gamma^m_{nq}+
\Gamma^i_{nq}\Gamma^m_{ip}-[p\leftrightarrow q].
\end{equation}
Therefore,
\begin{eqnarray}
R_{mpqn} &=& g_{mq}g_{np}-g_{pq}g_{mn} \;, \label{curv2} \\
\nabla_m R_{pqrs} &=& 0   \label{Rvanish},
\end{eqnarray}
where $\nabla_m$ is the usual covariant derivative on the manifold.

For simplicity, let us introduce the
notation 
\begin{equation}
  G_{\mu\nu}(\pi) \equiv g_{mn}\partial_{\mu}\pi^{m}\partial_{\nu}\pi^{n},
\end{equation}
and we use $[G_{\mu\nu}(\pi)]_{\xi^n}$
to denote the $n$th term of the Taylor expansion of the operator
 $G_{\mu\nu}(\pi)$
in terms of the normal coordinates $\xi$.
With this notation, the  first 
four terms of the normal coordinate expansion  are
\begin{eqnarray}
[G_{\mu\nu}(\pi)]_{\xi^1} & = &
    g_{mn}(\psi)D_{\mu}\psi^{m}\partial_{\nu}\xi^{n} + (\mu \leftrightarrow \nu),
\label{Exp1} \\
\left[G_{\mu\nu}(\pi)\right]_{\xi^2}
 & = &
    g_{mn}(\psi)D_{\mu}\xi^{m}D_{\nu}\xi^{n} + 
    R_{mijn}(\psi)\partial_{\mu}\psi^{m}\partial_{\nu}\psi^{n}\xi^{i}\xi^{j},
\label{Exp2} \\
\left[G_{\mu\nu}(\pi)\right]_{\xi^3} & = &
    \frac{1}{6}R_{mijk}(\psi)\partial_{\mu}\psi^{m}D_{\nu}\xi^{k}\xi^{i}\xi^{j} + 
     (\mu \leftrightarrow \nu),
           \label{Exp3}    \\ 
\left[G_{\mu\nu}(\pi)\right]_{\xi^4} & = &
  \frac{1}{6}R^{p}_{ijm}(\psi)R_{pkln}(\psi)
\partial_{\mu}\psi^{m}\partial_{\nu}\psi^{n} +
   R_{mijn}D_{\mu}\xi^{m}D_{\nu}\xi^{n}\xi^{i}\xi^{j} \label{Exp4};  
\end{eqnarray}
 where  
the covariant derivative $D_{\mu}$ is defined by
\begin{equation}
D_{\mu}\xi^{m} \equiv \partial_{\mu}\xi^{m}+
\Gamma_{st}^{m}\partial_{\mu}\psi^{t}\xi^{s}, \label{Covar1}
\end{equation}
and we have used Eq. (\ref{Rvanish}).
The expansion of an invariant quantity 
in terms of the normal coordinates 
is manifestly reparametrization invariant: the coefficients
that multiply the monomials $\xi^m (x)$ are tensors. Note that these coefficients, the
curvature tensor and the metric, depend on the classical field $\psi(x)$ and 
are not affected by the integration over the quantum field,
$\xi$.
 
The divergences that determine the renormalization constants
$Z_{ij}$ are obtained  by integrating over the quantum 
field $\xi$. More specifically, we calculate the
one-particle irreducible diagrams of the 
expectation value
\begin{equation}
\langle O^{(e)}(\psi,\xi) \rangle   =
                    \frac{\int[d\xi]O^{(e)}(\psi,\xi)e^{-S^{(e)}(\psi,\xi)}}
                        {\int[d\xi]e^{-S^{(e)}(\psi,\xi)}}, 
\label{Expectation}
\end{equation}
where $O^{(e)}$ and $S^{(e)}$ stand for
all the terms of $O(\xi^{n})$ with $n\geq2$ of 
the expansion in terms of the normal coordinates. The expansion of $e^{-S^{(e)}}$
in powers of $\xi$  
and the integration $[d \xi]$ generates the diagramatic expansion.

In order to calculate the Feynmann diagrams generated from Eq. (\ref{Expectation}),
it is necessary to compute the
propagator of the field $\xi$. However, when we use Eq. (\ref{Exp2}) and Eq.
(\ref{Covar1}) to obtain the noninteracting part of the action, we find that
\begin{eqnarray}
[S]_{\xi^2} & = & \frac{1}{2}\int dx \left \{ 
                     g_{mn} D_{\mu}\xi^{m} D^{\mu}\xi^{n}
                     + R_{iklj}\partial_{mu} \psi^{i}
                     \partial^{mu} \psi^{j}
                    \xi^{k} \xi^{l} \right \}
                                              \nonumber \\
& = &    \frac{1}{2}\int dx \left \{g_{mn}
  \partial_{\mu}\xi^{m}
  \partial_{\mu}\xi^{n} + \ldots
\right \}  \label{Quadra}.
\end{eqnarray}
This leads to a complicated propagator that depends on the
metric. Although it is possible to continue with the  calculation, 
it is simpler to perform another
tranformation of coordinates to obtain the more common  propagator.
The transformation
is between the
curved system of coordinates to the tangent
system of coordinates: 
\begin{eqnarray}
\xi^{a}&=&e^{a}_{i}(\psi)\xi^{i}, \\
\xi^{i}&=&e^{i}_{a}\xi^{a}, \\
e^{ai}(\psi)g_{ij}e^{bj} (\psi)&=&\delta^{ab} \label{Vielbein}
\end{eqnarray}
where the local matrix $e^{a}_{m}$ is known as the {\em Vielbein}.
Here, we follow the convention of using the earlier letters of the
latin alphabet (a, b, etc.) for the local indices
and the latter indices (i, j, etc.) for the covariant indices. Furthermore, 
$\delta_{ab}$ is the diagonal matrix with the diagonal elements
$(1,1,\ldots,1)$, so there is no distinction between covariant and 
contravariant indices.

In terms of
these local coordinates, the covariant derivative becomes
\begin{eqnarray}
D_{\mu}\xi^{a} & \equiv & e^{a}_{m}D_{\mu}\xi^{m}, \nonumber\\
  &=& \partial_{\mu}\xi^{a}+A(\psi)_{\mu}^{ab} \psi^{i}\xi^{b}
 \label{Covar2};
\end{eqnarray}
where the quantity $A(\psi)_{\mu}^{ab}$ has dimension unity and
transforms as a gauge field
under the rotations of the 
tangent frames defined by  $e^{a}_{i}$.
So, the only way in which $A(\psi)_{\mu}^{ab}$
appears in the calculation is through the
field strength $F_{\mu\nu}^{ab}$, which is
\begin{eqnarray}
F_{\mu\nu}^{ab}  &=  &\partial_{\mu}A^{ab}_{\nu}-\partial_{\nu}A^{ab}_{\mu}+
        A^{ac}_{\mu}A^{cb}_{\nu}-A^{ac}_{\nu}A^{cb}_{\mu} \nonumber \\
        &  =     &
e^{ai}e^{bj}R_{ijmn}\partial_{\mu}{\psi}^{m}
\partial_{\nu}{\psi}^{n} . \label{identity}
\end{eqnarray}

The substitution of the local coordinates $\xi^{a}$ in 
Eq. (\ref{Quadra}) yields
\begin{eqnarray}
[S]_{\xi^2} & = & \frac{1}{2}\int dx \left \{ D_{\mu}\xi^{a}D_{\mu}\xi^{a}+
 e^{i}_{a} e^{j}_{b} R_{mijn}
\partial_{\mu}\psi^{m}\partial_{\mu}\psi^{n}\xi^{a}\xi^{b} 
                               \right \}
                                 \nonumber \\
& = &    \frac{1}{2}\int dx \left \{
 (  \partial_{\mu}\xi^{a}+ A_{\mu}^{ab}\xi^{b}  ) 
(   \partial_{\mu}\xi^{a}+ A_{\mu}^{ab}\xi^{b}  ) +
 R_{mabn}\partial_{\mu}\psi^{m}\partial_{\mu}\psi^{n}\xi^{a}\xi^{b} 
\right \}  \label{Quadra2}
\end{eqnarray}
from which we  obtain the usual propagator
\begin{equation}
\langle\xi^{a}(x) \, \xi^{b}(y)\rangle = 
                \delta^{ab}\int dp\frac{e^{ip(x-y)}}{p^2+m^2},
\end{equation}
In this equation $m^{2}$ is an infrared cutoff.

\section{The One Loop Calculation}
We now compute the one-loop correction to demonstrate how the background field method
works and to reproduce the results obtained by Wegner\cite{Wegner1}.
A simple rescaling of the field $\xi$ 
shows that, to  one-loop order, we only need to
expand the action and the operator up to order O($\xi^2$). 
Using the notation introduced in the previous section,
we have to compute the one-particle irreducible diagrams corresponding to 
the expectation value 
\begin{equation}
\langle \left[G^{s}_{\mu\nu}(\pi)\right]_{\xi^2}\rangle   =
 \frac{
\int[d\xi] \left[ G_{\mu\nu}^{s}(\pi) \right]_{\xi^2}
  e^{ \left[ S(\pi) \right]_{\xi^2}}
            }
      {
\int[d\xi]  e^{ \left[ S(\pi) \right]_{\xi^2}}
                  }
\label{Expect2}
\end{equation}
Consider the simplest case $s=2$.
Then
\begin{eqnarray}
[G^2_{\mu\mu}(\pi)]_{\xi^2} &=&
G_{\mu\mu}(\psi)[G_{\alpha\alpha}(\pi)]_{\xi^2}+ 
[G_{\mu\mu}(\pi)]_{\xi^1}  [G_{\alpha\alpha}(\pi)]_{\xi^1} +\nonumber \\
& & \;\;\;\;\;\; + [G_{\mu\mu}(\pi)]_{\xi^2} G_{\alpha\alpha}(\psi),
\end{eqnarray}
where the expression for $[G_{\mu\mu}]_{\xi^2}$ 
is given in Eq. (\ref{Exp2}) and, the term
$[G_{\mu\mu}(\pi)]_{\xi^1}
[G_{\alpha\alpha}(\pi)]_{\xi^1}$
can be found  from Eq. (\ref{Exp1}):
\begin{eqnarray}
[G_{\mu\nu}(\pi)]_{\xi^1}
[G_{\alpha\beta}(\pi)]_{\xi^1} &=&
e_{mc}e_{ia}\partial_{\mu}\psi^{m}\partial_{\alpha}\psi^{i}D_{\nu}\xi^{c}
D_{\alpha}\
\xi^{a} \nonumber\\
&+&(\alpha \leftrightarrow \beta)+
 (\mu \leftrightarrow \nu)\label{exp11} \\
 &+&    (\alpha \leftrightarrow \beta,\mu \leftrightarrow \nu) .\nonumber
\end{eqnarray}
Note that the metric has been written in terms of the {\em Vielbeins}.
Substituing in  Eq. (\ref{Expect2}) and remembering that $\psi(x)$
is a classical field, we get
\begin{eqnarray}
\langle \left[G^{2}_{\mu\nu}(\pi)\right]_{\xi^2}\rangle  & =  &
G_{\mu\mu}(\psi)
\langle [G_{\alpha\alpha}(\pi)]_{\xi^2} \rangle +
G_{\mu\mu}(\psi)
\langle [G_{\alpha\alpha}(\pi)]_{\xi^2} \rangle     \nonumber  \\
&+& \langle [G_{\mu\mu}(\pi)]_{\xi^1}  [G_{\alpha\alpha}(\pi)]_{\xi^1}\rangle.
\end{eqnarray}
The extraction of the divergences
of the one-loop 
diagrams is carried out with the help of the dimensional regularization;
the details are given in Appendix A. We find 
 \begin{eqnarray}
{\langle [G_{\mu\mu}(\pi)]_{\xi^2}\rangle}_{W}
 &=&
\frac{-1}{2\pi\epsilon}  R_{mn} [ \partial_{\mu}\psi^{m}\partial_{\nu}\psi^{n}
                             -\frac{\delta_{\mu\nu}}{2} 
                           \partial_{\mu}\pi^{m}\partial^{\nu}\pi^{n}],
             \label{rs1.1a}\\
{\langle[G_{\mu\nu}(\pi)]_{\xi^1}
[G_{\alpha\beta}(\pi)]_{\xi^1}\rangle}_{W}   &=&
\frac{1}{4\pi\epsilon} R_{mpqn} [ \delta_{\mu\alpha} 
\partial_{\gamma}\psi^{m}\partial^{\gamma}\psi^{n}
\partial_{\nu}\psi^{p}\partial_{\beta}\psi^{q} +  \nonumber \\ 
& & \;\;\;\;\;\;\;\;\;\;
\partial_{\mu}\psi^{m}\partial_{\nu}\psi^{n}
\partial_{\alpha}\psi^{p}\partial_{\beta}\psi^{q} ] +   \nonumber \\
& & (\mu \leftrightarrow \nu) + (\alpha \leftrightarrow \beta)+
(\mu \leftrightarrow \nu  ,  \alpha \leftrightarrow \beta). \label{rs1.2a}
\end{eqnarray}
For the manifold $O(N)/O(N-1)$, we can substitute the expression for   
$R_{mnpq}$ given earlier and obtain
 \begin{eqnarray}
{\langle[G_{\mu\nu}(\pi)]_{\xi^2}\rangle}_{W}
 &=&
\frac{N-2}{2\pi\epsilon} [G_{\mu\nu}(\psi) -  \frac{\delta_{\mu\nu}}{2}
G_{\alpha\alpha}(\psi)],    \label{rs1.1b}  \\
\langle [G_{\mu\nu}(\pi)]_{\xi^1}
[G_{\alpha\beta}(\pi)]_{\xi^1}\rangle_{W}                                       
                    &=&
        2G_{\mu\nu}(\psi)G_{\alpha\beta}(\psi)  -
 \left [G_{\mu\alpha}(\psi)G_{\nu\beta}(\psi)  +
                 (\mu \leftrightarrow \nu) \right ]\nonumber\\
&+& \left \{   \frac{\delta_{\mu\alpha}}{2}   \left [
(G_{\gamma\gamma}(\psi)G_{\nu\beta}  
 -G_{\nu\gamma}(\psi)G_{\beta\gamma}  \right  ]   \right \}\nonumber\\
&+& 
   \{ \alpha \leftrightarrow\beta \} + 
   \{\mu \leftrightarrow \nu  \} +  
   \{\alpha \leftrightarrow\beta , \mu \leftrightarrow \nu \}\label{rs1.2b}.
\end{eqnarray}
These equations are identical to
those obtained by Wegner\cite{Wegner1}. Note
that the operator 
$G^{2}_{\mu\nu}(\pi)$
mixes with the operator
$G_{\mu\beta}G_{\beta\nu}$. Furthermore, it also follows that 
the  operators that mix with 
$G_{\mu_{1}\nu_{1}}(\pi) \ldots G_{\mu_{s}\nu_{s}}(\pi)$
are the cyclic products of the  form
\begin{equation}
O^{\rm cyc}\equiv(g_{ij}\partial_{\alpha}\pi^{i}\partial_{\beta}\pi^{i})  
(g_{mn}\partial_{\beta}\pi^{m}\partial_{\rho}\pi^{n})  \ldots
(g_{kl}\partial_{\gamma}\pi^{k}\partial_{\alpha}\pi^{l}) \label{cyclic}
\end{equation}
This is expected since within the dimensional regularization scheme,
 the operators that mix have the same symmetry and the same dimension.
Surprisingly, the operators
\begin{equation}
(g_{ij} 
(\partial_{\mu_1})^{m_1} \pi^{i} (\partial_{\mu_2})^{m_2} \pi^{j})
(g_{kl}
(\partial_{\mu_3})^{m_3} \pi^{k} (\partial_{\mu_4})^{m_4} \pi^{l})
\ldots
\label{mixother},
\end{equation}
with $m_1+m_2+m_3+ \cdots=2s$ do not mix, 
even though they have the same
dimension, and are also invariant. 

We shall be studying the renormalization of the operators $O^{\rm cyc}$. 
Let us imagine that there are $N$ such operators denoted by the set $\{O^{\rm
cyc}_i\}$. We shall denote all other operators by $P$, and let there be
$M$ such operators.
% We also note  two general rules: (a) a composite
%operator of dimension $y$ mixes with operators of dimension $y'\le y$; (b) within
%the dimensional regularization scheme, the operators that mix have the same
%symmetry and the same dimension.
 Then, the renormalized operators can be
expressed as 
\begin{eqnarray}
O^R_i&=&\sum_{j=1}^NC_{ij}O_j+\sum_{j=1}^ME_{ij}P_j, \nonumber \\
P^R_i&=&\sum_{j=1}^NF_{ij}O_j+\sum_{j=1}^MD_{ij}P_j.
\end{eqnarray}
The one-loop  matrix $Z^{(1)}_{ij}$
for the invariant operators with
$2s$ gradients will have the form 
\begin{equation}
Z^{(1)}(\epsilon,t)= \left[
\begin{array}{ll}
     C^{(1)}&  0\\
     0& D^{(1)}  
\end{array}
\right].
\end{equation}
Wegner \cite{Wegner1} has shown that 
the largest eigenvalue comes from the matrix $C^{(1)}$, and consequently we shall
not concern ourselves with  the matrix $D^{(1)}$.

Although  we know the
form of the operators that mix under renormalization, it is still
a difficult task to diagonalize the matrix $C^{(1)}_{ij}$
since we do not have a simple way of clasifying the
operators. Near
two dimensions the problem can be solved if we introduce conformal
coordinates\cite{Kravstov}, so that
\begin{equation}
\partial_{+}=\partial_{x}+i\partial_{y},\,\,\,
 \partial_{-}=\partial_{x}-i\partial_{y} \label{compder}.
\end{equation}
In terms of these new coordinates, the divergences in 
 Eq. (\ref{rs1.1b}) and Eq. (\ref{rs1.2b})
take  very simple forms:
 \begin{eqnarray}
\langle [G_{++}(\pi)]_{\xi^2}\rangle
&=&\frac{(N-2)}{2\pi\epsilon}G_{++}(\psi),
\label{rs1.1c}\\
\langle [G_{--}(\pi)]_{\xi^2}\rangle
&=&\frac{(N-2)}{2\pi\epsilon}G_{--}(\psi),\\
\langle [G_{+-}(\pi)]_{\xi^1}
[G_{+-}(\pi)]_{\xi^1}\rangle  &=&
\frac{-1}{\pi\epsilon}
[G_{+-}(\psi)^{2} - 
G_{++}(\psi)G_{--}(\psi)]\label{rs1.2c},
\end{eqnarray} 
while all other possibilities vanish.
A vanishing result simply means that the expectation value is 
free from divergences.

Let us now introduce the following notations:
\begin{eqnarray}
H&\equiv& g_{kl}\partial_{+}\pi^k\partial_{-}\pi^l, \\
A&\equiv& g_{kl}\partial_{+}\pi^k\partial_{+}\pi^l,\\
B&\equiv&g_{kl}\partial_{-}\pi^k\partial_{-}\pi^l.
\end{eqnarray}
{\it We now derive the important result that the operator $H^s$ mixes only with
the set of operators}
\begin{displaymath}
\left\{H^s, H^{s-2}(AB), H^{s-4}(AB)^2, \cdots , H^2(AB)^{{s\over 2}-1},
(AB)^{s\over 2}\right\}.
\end{displaymath}
Let us denote the normal coordinate expansion of an operator by
\begin{equation}
Q(\pi)=Q(\psi)+q_1(\psi,\xi)+q_2(\psi,\xi)+\cdots
\end{equation}
and the one-loop contribution by
\begin{equation}
\langle Q(\pi)\rangle_{1L}=\langle q_2(\psi,\xi)\rangle.
\end{equation}
The last equation follows because only the quadratic term contributes to 
one-loop order, and the  average implies an integration over the normal
coordinates. From Eqs. (\ref{rs1.1c}-\ref{rs1.2c}), it follows that
\begin{eqnarray}
\langle a_2(\psi,\xi)\rangle_{1L}&=&\nu I A(\psi),\label{I}\\
\langle b_2(\psi,\xi)\rangle_{1L}&=&\nu I B(\psi),\label{II}\\
\langle h_1(\psi,\xi)
h_1(\psi,\xi)\rangle_{1L}&=&-2I\left[H^2(\psi)-A(\psi)B(\psi)\label{III}
\right],
\end{eqnarray}
where $\nu=(N-2)$ and $I=1/2\pi\epsilon$.

The one-loop computation, using Eqs. (\ref{I}-\ref{III}), shows that
\begin{equation}
\langle H^s\rangle_{1L}=-s(s-1)I\left[H^s(\psi)-H^{s-2}A(\psi)B(\psi)\right].
\end{equation}
Similarly,
\begin{equation}
\langle H^{s-2}AB\rangle_{1L}=IH^{s-2}(\psi)A(\psi)B(\psi)\left[(s-1)(s-2)-2\nu\right]
-I(s-1)(s-2)H^{s-4}(\psi)\left[A(\psi)B(\psi)\right]^2.
\end{equation}
Recursively, one can show that $H^s$ mixes only with the operators listed above. It is now easy to show that 
\begin{eqnarray}
\langle (AB)^lH^{s-2l}\rangle_{1L}=&-&IH^{s-2l}(\psi)\left[A(\psi)B(\psi)\right]^l\left[(s-2l)(s-2l-1)-2\nu l\right]\nonumber\\
&+&I(s-2l)(s-2l-1)H^{s-2l-2}(\psi)\left[A(\psi)B(\psi)\right]^{l+1}\label{oneloop}.
\end{eqnarray}
From these
equations we can now derive the renormalization matrix $Z^{(1)}(t,\epsilon)$.

Using Eq. (\ref{oneloop}), 
we can read off the matrix $Z_{ij}(t,\epsilon)$, which 
is a $[\frac{s}{2}]\times[\frac{s}{2}]$ matrix:
\[ Z^{(1)}(\epsilon,t)-1= \frac{t}{2\pi\epsilon}\left[
\begin{array}{lllll}
a_{11}&     a_{12}&             0&              0&    \cdots \\
     0&      a_{22}&           a_{23}&          0&     \cdots \\
     0&         0&             a_{33}&          a_{34}&    0\\
     0&          0&               0&           \vdots&     \vdots \\
 \vdots&   \vdots&             \vdots&           \vdots&     \ddots 
\end{array}
\right], \]
where 
\begin{eqnarray}
a_{j+1 j+1}&=& (s-2j)(s-2j-1)-2\nu j , \\
a_{j+1 j+2}&=& -(s-2j)(s-2j-1),
\end{eqnarray}
$j=0,1,2\ldots,[\frac{s}{2}]$. Note that there is a change in sign because 
to calculate $Z$ we have to subtract a divergence. Because this matrix is upper
triangular, the diagonal elements are the eigenvalues. The largest eigenvalue is
the element $a_{11}$. Using Eq. (\ref{anomal}) and (\ref{yalpha}), we get 
\begin{eqnarray}
y_{2s}^{(1)}&=& d-2s-\gamma^{(1)}(\epsilon)\\
           & =& 2+\epsilon\left(1+\frac{s(s-1)}{N-2}\right),
\end{eqnarray}
where we have substituted the well-known one-loop value of the fixed point
$t^*=2\pi\epsilon/\nu$. Therefore, for sufficiently large $s$, regardless of 
how small $\epsilon$ is, an infinite number of  high gradient operators become
relevant.

For the two-loop calculation we shall also need the eigenvectors of the matrix
$C^{(1)}$. They are compactly contained in the matrix $S$ that
diagonalizes $C^{(1)}$. The  matrix $S$ is
\begin{equation}
S=\left(
\begin{array}{ccccc}
1&         x_{12}&         x_{13}&         x_{14}&        \cdots    \\
0&    1&         x_{23}&         x_{24}&         \cdots \\
0&    0&    1&         x_{34}&         \cdots   \\
0&    0&    0&    1&         \cdots   \\
\vdots&   \vdots&             \vdots&           \vdots&     \ddots  
\end{array}
\right).
\end{equation}
where
\begin{equation}
x_{ii+l}=\frac{a_{i i+1}a_{i+1 i+2}\cdots a_{i+l-1 i+l}}
{(a_{i i}-a_{i+1 i+1})(a_{i i}-a_{i+2 i+2})\cdots (a_{i i}-a_{i+l i+l})} 
\end{equation}
It is easily verified that $S C^{(1)}S^{-1}$ is diagonal and the
diagonal elements are the diagonal elements of the matrix $C^{(1)}$. 
 
\section{The Two Loop Calculation}
 
The two-loop calculation is more involved. We need to  subtract the subdivergences
from the two-loop diagrams and to regulate the infrared divergences.
The method  for subtracting divergences that we 
follow is due to Bogoliubov-Parasiuk-Hepp-Zimmerman (BPHZ). This method
consists of subtracting directly the subdivergences from each Feynamnn diagram by
using the forest formula of Zimmerman\cite{Zimm}. For a complete discussion of
the forest formula and detailed examples, we refer the reader to
Collins\cite{Coll84}. With respect to infrared divergences, their
presence is due to the absence of
a mass term in the action. Infrared divergences make
the computation of the loop integrals ambiguous when dimensional regularization is used. The poles due to the ultraviolet
and the infrared divergences tend to cancel each other, leading in some cases
to a vanishing result. The easiest way to solve this problem is to 
introduce an infrared cutoff in every propagator,
{\it i.e.}, we let $p^2 \rightarrow p^2+m^2$
in the internal lines of a diagram. 
Of course,  other choices of cutoff should 
not affect the final result since the dependence on the cutoff appears only
in  terms  corresponding to subdivergences. These terms are
eventually subtracted
with the forest formula.

As with  the one-loop calculation, we begin
by considering the normal coordinate expansion of the operator
$(G_{\mu\nu})^s$ and of the action. From  a simple rescaling of the fields we
learn that the expansion in terms of the normal coordinate needs to be carried
out to order
$O(\xi^4)$. This time, a calculation of the divergences for the
case $s=4$  is sufficient to determine the divergences  
for arbitrary $s$. The different possibilities that arise in the
expansion of $(G_{\mu\nu})^4$ are:
\begin{eqnarray}
O(\xi^2)&:&[G_{\mu\nu}(\pi)]_{\xi^2}\; ,
         \, [G_{\mu\nu}(\pi)]_{\xi^1}
          [G_{\alpha\beta}(\pi)]_{\xi^1} 
                           \label{xi2};        \\
O(\xi^3)&:&[G_{\mu\nu}(\pi)]_{\xi^3}\; ,
         \,[G_{\mu\nu}(\pi)]_{\xi^2}
          [G_{\alpha\beta}(\pi)]_{\xi^1}\; , \nonumber              \\
& &    [G_{\mu\nu}(\pi)]_{\xi^1}
         [G_{\alpha\beta}(\pi)]_{\xi^1}
          [G_{\eta\rho}]_{\xi^1} ; \label{xi3}   \\
O(\xi^4)&:& [G_{\mu\nu}(\pi)]_{\xi^4}\; ,
         \,[G_{\mu\nu}(\pi)]_{\xi^2}
          [G_{\alpha\beta}(\pi)]_{\xi^2}\; , \nonumber              \\
& &    [G_{\mu\nu}(\pi)]_{\xi^1}
         [G_{\alpha\beta}(\pi)]_{\xi^1}
          [G_{\eta\rho}]_{\xi^2}\;\;,
        [G_{\mu\nu}(\pi)]_{\xi^1}
         [G_{\alpha\beta}(\pi)]_{\xi^1}
          [G_{\eta\rho}]_{\xi^1} [G_{\eta\rho}]_{\xi^1}. \label{xi4}
\end{eqnarray}
Except for the possibility of an arbitrary
number of two-point insertions in the internal lines,
\begin{figure}[htb]
\centerline{\epsfxsize 9cm
\epsffile{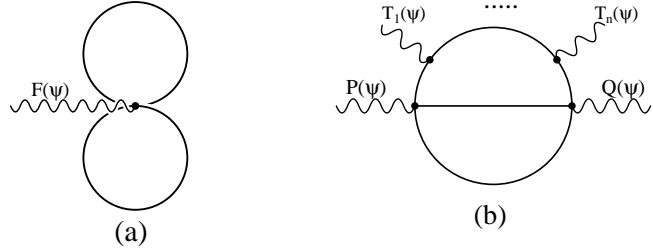}
}
\caption{The general form  of the two-loop diagrams. 
The wiggly lines represent functions that depend on 
the background field $\psi$, which  is  a
classical field.
Diagrams of type (a) lead to contributions proportional to
double poles in $\epsilon$ and are therefore not necessary for the computation of  anomalous dimension. Only graphs of type (b) need to be considered.}
\label{twoloop1}
\end{figure}
Fig. ~\ref{twoloop1} shows schematically the  two  types of
diagrams that arise in the two-loop calculation.
In Appendix B it is shown  that the graphs of Fig. ~\ref{twoloop1}(a) are
either finite or lead to double poles in $\epsilon$
after  the forest formula is applied to them.
 Hence, this type of  graph  plays no role in  determining
critical exponents. Moreover, the term of order $O(\xi^4)$ appearing in the
expansion of  $(G_{\mu\nu})^s$, and in the action, can be discarded  because the
two-loop diagrams generated by the 4-point vertices are of the type shown in Fig.
~\ref{twoloop1}(a). In view of this result,  the two-loop
calculation 
requires only to carry out the expansion of an operator $Q(\pi)$
in terms of normal coordinates to order $O(\xi^3)$,
\begin{equation}
Q(\pi)=Q(\psi)+q_1(\psi,\xi)+q_2(\psi,\xi)+q_3(\psi,\xi)+\cdots,
\end{equation}
and the {\it only} expectation
values needed are: 
\begin{eqnarray}
O(\xi^2):& &
\langle [G_{\mu\nu}]_{\xi^2}\rangle,  \;
\langle [G_{\mu\nu}]_{\xi^1}
                 [G_{\alpha\beta}]_{\xi^1}\rangle 
             \label{twoloopexp1} \\
O(\xi^3):& &
\langle [G_{\mu\nu}]_{\xi^1}[G_{\alpha\beta}]_{\xi^2}\rangle ,\;
\langle [G_{\mu\nu}]_{\xi^1}[G_{\alpha\beta}]_{\xi^1}
            [G_{\kappa\lambda}]_{\xi^1}\rangle    \label{twoloopexp2}
\end{eqnarray}
In principle, it is also necessary to calculate the divergences
of the operators in Eq. (\ref{mixother}) as they may mix
with the operators we are considering. However, we shall show that this
mixing  may be neglected without  affecting our
conclusions. 
A detailed calculation of the expectation value
$
\langle [G_{\mu_1\nu_1}]_{\xi^1}[G_{\mu_2\nu_2}]_{\xi^2} \rangle
$
is given in Appendix B. 

We emphasize again that we shall study only the renormalization of the set of
operators
\begin{displaymath}
\left\{H^s, H^{s-2}(AB), H^{s-4}(AB)^2, \cdots , H^2(AB)^{{s\over 2}-1},
(AB)^{s\over 2}\right\}.
\end{displaymath}
The calculation is tedious, but the essential pieces are given in Appendix D. We
find that 
\begin{eqnarray}
\langle H^s\rangle_{2L}=&-&{H^s(\psi)\over \omega}\left[3\nu s +{s(s-1)\over
2}(\nu+6)+2s(s-1)(s-2)\right]\nonumber \\
&+&{H^{s-2}(\psi)A(\psi)B(\psi)\over \omega}\left[{s(s-1)\over
2}(3-2\nu)+2s(s-1)(s-2)\right],
\end{eqnarray}
where $\omega=24\pi^2\epsilon$. Similarly,
\begin{eqnarray}
\langle H^{s-2}AB\rangle_{2L}&=&{H^s(\psi)\over
\omega}\left[9-\nu+14(s-2)\right]\nonumber \\
&+&{H^{s-2}(\psi)A(\psi)B(\psi)\over \omega}\Bigg[(7\nu
-9)-(s-2)(9\nu+14)\nonumber \\
&-&{1\over
2}(s-2)(s-3)(\nu+2)-2(s-2)(s-3)(s-4)\Bigg]\nonumber \\
&+&{H^{s-4}(\psi)\left(A(\psi)B(\psi)\right)^2\over \omega}
\Bigg[{1\over 2}(s-2)(s-3)(2-{13\nu\over 2})\nonumber \\ 
&+&2(s-2)(s-3)(s-4)\Bigg].
\end{eqnarray}
The general case is even more tedious, but the necessary ingredients are given
in Appendix D. We find that
\begin{eqnarray}
\langle
H^{s-2l}(AB)^l\rangle_{2L}&=&r'_{l-1l}H^{s-2l+2}(\psi)(A(\psi)B(\psi))^{l-1}
+r'_{ll}H^{s-2l}(\psi)(A(\psi)B(\psi))^{l}\nonumber \\
&+&r'_{ll+1}H^{s-2l-2}(\psi)(A(\psi)B(\psi))^{l+1}+O({1\over
\epsilon^2})+{\rm other\  operators},
\end{eqnarray}
where
\begin{eqnarray}
r'_{ll}&=&-{1\over \omega}\Bigg[2(s-2l)(s-2l-1)(s-2l-2)+14l^2(s-2l)\nonumber \\
&+&2l(s-2l)(s-2l-1)+4l^2(l-1)+{1\over 2}(s-2l)(s-2l-1)(\nu+6)+6\nu
l(s-2l)\nonumber \\
&-&l^2(\nu-9)-6\nu l-3\nu l (l-1)\Bigg]\label{melement1},\\
r'_{l,l+1}&=&{1\over
\omega}\Bigg[2(s-2l)(s-2l-1)(s-2l-2)-2l(s-2l)(s-2l-1),\nonumber \\
&+&{1\over 2}(s-2l)(s-2l-1)(s-{13\over 2}\nu)\Bigg]\label{melement2}\\
r'_{l,l-1}&=&{1\over \omega}\Bigg[l^2(9-\nu)+4l^2(s-2l)+4l^2(l-1)\Bigg]\label{melement3}.
\end{eqnarray}

The matrix $\gamma$ in Eq. (\ref{anomal}) can now be written in the  form
\begin{equation}
\gamma(t,\epsilon)= -at-2bt^2+O(t^3)\label{RNG}
\end{equation}
where the matrices $a$ and $b$ are
\begin{equation}
\begin{array}{c}
a=\left(
\begin{array}{cc}
C^{(1)}&  0  \\
 0     & D^{(1)} 
\end{array}
\right) ,\ 
b=\left(
\begin{array}{cc}
C^{(2)}&  X \\
 Y     & D^{(2)} 
\end{array}
\right)
\end{array} 
\vspace{.2cm}
.  
\end{equation}
The matrix $C^{(1)}$ was obtained in the previous section. 
From Eqs. (\ref{melement1}-\ref{melement3}) we find that the matrix $C^{(2)}$ is
\begin{equation}
  C^{(2)}=  \frac{1}{12\pi^2}\left[
\begin{array}{llllll}
r_{11}&     r_{12}&            0&         0&      0&  \cdots \\
 r_{21}&    r_{22}&       r_{23}&         0&       0&  \cdots \\
     0&     r_{32}&        r_{33}&      r_{34}&   0&      \cdots \\
     0&          0&         r_{43}&       r_{44}&   r_{45}&  \cdots \\
 \vdots&   \vdots&       \vdots&        \vdots&     \vdots&  \ddots \\
 \vdots&     \vdots&        \vdots&       \vdots&    \vdots& \vdots 
\end{array}
\right] 
\end{equation}
where 
\begin{eqnarray}
r_{j+1\,j+1} &=& -\omega r_{j\,j}^{'}, \\
r_{j+1\,j+2} &=& -\omega r_{j\,j+1}^{'}, \\
r_{j+1\,j} &=& -\omega r_{j-1\,j}^{'},
\end{eqnarray}
%%
%%
%%
%%
%%
%\begin{eqnarray}
%r_{j+1\,j+1} &=& 2(s-2j)(s-2j-1)(s-2j-2)
%+\frac{1}{2}(\nu +6)(s-2j)(s-2j-1) +\nonumber \\
%& & 3 \nu (s-2j)-2j(s-2j)(s-2j-1)+14j^2(s-2j)+6 \nu j(s-2j) +
%                \nonumber\\
%& & -6 \nu j- 3\nu j(j-1)-j^2(\nu -9)+ 4j^2(j-1), \vspace{.5cm}\\
%r_{j+1\,j+2} &=& -2(s-2j)(s-2j-1)(s-2j-2)+  \nonumber \\
%   &  &-  \frac{1}{2}(6-\frac{13\nu}{2})(s-2j)(s-2j-1)+
% 2j(s-2j)(s-2j-1), \vspace{.5cm}
%\\
%r_{j+1\,j} &=& -14j^2(s-2j)-4j^2(j-1)+j^2(\nu -9)\;
%,
%\end{eqnarray}
and $j=0,1,...[\frac{s}{2}]$, $\nu \equiv N-2$.

To calculate the anomalous dimension correct to two-loop
order, it is sufficient to use elementary first order nondegenerate
perturbation theory since the eigenvalues of $C^{(1)}$  are nondegenarate.
Let $U$ be the transformation matrix that diagonalizes the one loop matrix, where
\begin{equation}
U=\left(\begin{array}{cc}
S & 0  \\
0 & T \end{array}\right).
\end{equation}
The matrix $S$ was given in the previous section. Therefore,
\begin{equation}
U\gamma U^{-1}=-t\left(\begin{array}{cc}
SC^{(1)}S^{-1} & 0 \\
0 &TD^{(1)}T^{-1} \end{array}\right)-2t^2
\left(\begin{array}{cc}
SC^{(2)}S^{-1} & SXT^{-1} \\
S^{-1}Y^{(1)}T & TD^{(2)}T^{-1} \end{array}\right).
\end{equation}
The two-loop correction to the anomalous dimension $y_{2s}$ is contained in the
diagonal entries of the second matrix. This is the reason why we could ignore
calculating the mixing matrices $X$ and $Y$. We of course do not have a
rigorous proof that a level crossing  does not occur and  that we do not need
the block $D^{(2)}$. However, this is highly unlikely within
perturbation theory. In any case, we shall show that by ignoring this block we
already obtain an eigenvalue corresponding to a positive full dimension
$y_{2s}$. A larger eigenvalue can only make things worse, while a smaller
eigenvalue does not change our conclusions. It is easy to see that the we need
only the
$(11)$-element,
$W_{11}$, of the matrix
$SC^{(2)}S^{-1}$, where
\begin{equation}
W_{11}=-{1\over \omega}\left[r_{11}+x_{12}r_{21}\right],
\end{equation}
with the matrix elements given above.
 
Using the fixed point $t=t^*=\frac{2\pi\epsilon}{\nu}(1-\frac{\epsilon}
{\nu})$,
and choosing $s$ sufficiently large, we find that, to two-loop order,
the full dimension $y_{2s}$ is given by 
\begin{equation}
y_{2s}=d-2s+ \frac{\epsilon s^2}{N-2} \left[1+O({1\over s})\right]+
\left[\frac{\epsilon^{2}s^3}{(N-2)^2}\right]\left[{2\over 3}+
O({1\over s})\right]+O(\epsilon^3)\label{y2s}.
\end{equation}
This is the central result of our paper. It shows that for any $\epsilon$,
however small, we can always find an infinite number of high gradient operators
that have positive scale dimension. 
 
\section{Renormalization group flows}
The main result of this paper is Eq. (\ref{y2s}). For any $\epsilon$, however
small, there are an infinite number of high gradient operators with positive
anomalous dimension $y_{2s}$. In fact, the dimension is larger, larger the
power of the  gradient operator.
The two-loop contribution has not changed the picture obtained from the
one-loop calculation\cite{Wegner1}, but has compounded the problem because the
two-loop contribution is even larger than the one-loop contribution for
sufficiently large $s$. 

The most curious phenomenon is the lack of feedback of the high gradient operators
to the gradient operators of lower powers. It has been shown from a perturbative
argument\cite{Wegner1} that the gradient operators of power $2s$ contribute to
the renormalization of the operators of powers $(4s-2)$. Thus, the situation is
very different form the $\phi^4$ theory
around four dimensions. In that instance, the gaussian fixed point
becomes unstable below 4 dimensions because the operator $u \phi^4$ becomes
relevant. However, a new non-trivial fixed point can be found because of the
feedback of  this  operator to the renormalization of the coupling associated
with $\phi^2$ term.  Thus, in contrast to $\phi^4$ theory, it is not possible to
locate a new stable fixed point within the $(2+\epsilon)$-expansion as described
in the present paper.

\begin{figure}[htb]
\centerline{\epsfxsize 9cm
\epsffile{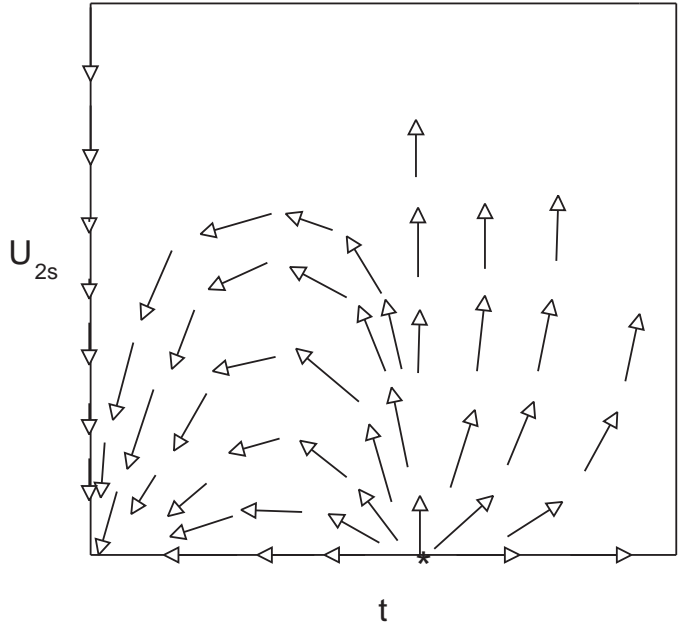}
}
\caption{The renormalization flows for $d=2+\epsilon$.
$U_{2s}$ is the charge associated with the high gradient operator
for sufficiently large $s$}
\label{rgflow.1}
\end{figure}

\begin{figure}[htb]
\centerline{\epsfxsize 9cm
\epsffile{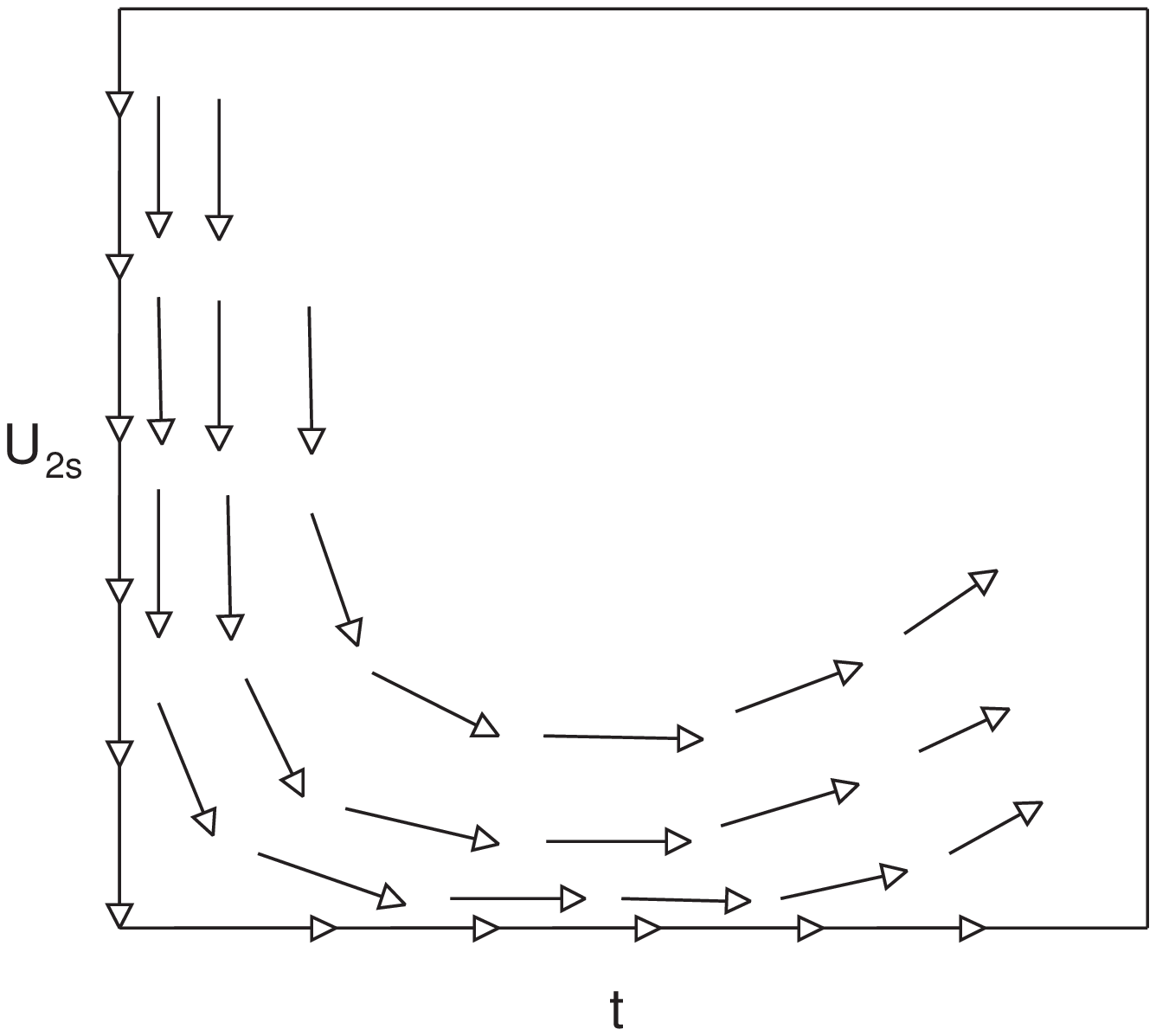}
}
\caption{The renormalization flows for $d=2$.
$U_{2s}$ is the charge associated with the high gradient operator
for sufficiently large $s$}
\label{rgflow.2}
\end{figure}

For a more complete understanding, consider the renormalization group equations.
From Eq. (\ref{RNG}), it is simple to see that
\begin{eqnarray}
\frac{dt}{dl}&=& -(d-2)t+(N-2)\frac{t^2}{2\pi}+
(N-2)\frac{t^3}{(2\pi)^2}+\cdots
            \label{flow1}     \\
\frac{1}{U_{2s}}\frac{dU_{2s}}{dl}  &= &
d-2s+\frac{ts(s-1)}{2\pi}+\frac{t^2[s^3+O(s^2)]}{12\pi^2} + \cdots
\label{flow2},
\end{eqnarray}
where $e^{l}$ is the rescaling factor, and
$U_{2s}$ is the coupling constant associated with an eigenoperator. 
The equation for $t$ is the well-known\cite{Heisenberg} equation. To recover the
previous result for
$y_{2s}$, it is only necessary to substitute the fixed point value of $t$
in the equation for $U_{2s}$. Written in this form, it is clear that the
difficulties persist for $d=2$; substitution of $\epsilon=0$ in Eq.
(\ref{y2s}) misleadingly leads one to believe that there are no difficulties in
$d=2$.
The renormalization group flows associated with these equations for
$d=2$ and $d=2+\epsilon$
are shown in Figs. ~\ref{rgflow.1} and ~\ref{rgflow.2}. 

Let us first consider the case $d=2+\epsilon$. In principle, for $t>t^*$,
there are two possibilities:
(1) $U_{2s} \to 0$,  and  (2) $U_{2s} \to \infty$.
The first possibility does not directly follow from these equations. This is
the hypothetical case in which the higher order terms, not calculated here, bend
the flows back.
Because the flows are vertical as they approach $t^*$ from below,
the magnetization must drop discontinuously at the transition\cite{Comm1}. Note
that due to the growth of the couplings $U_{2s}$ the spins cannot fluctuate from
the preferred direction because $U_{2s}$ multiplies the corresponding high
gradient operator. Therefore, it is energetically infinitely costly to allow the
gradient of the spins to be non-vanishing.  In the second possibility, in which 
$U_{2s} \to \infty$, an infinite number of
$U_{2s}$ grows simultaneously. This makes it exceedingly costly for the spins
in the system to point in different directions, regardless of $t$. In other
words, the spins cannot disorder. Non-perturbative effects, not contained, in
the $(2+\epsilon)$-expansion  are necessary to achieve a order-disorder
transition

The case $d=2$ is not much different.
The growth of $U_{2s}$
as $t$ increases leads us, once again, to one of the previous conclusions, 
that is, the system cannot disorder.   However,  there appears to be  a
region in which the $U_{2s}$ initially decreases with increasing length scale.
Therefore, for shorter length scales, the perturbative renormalization
group equations may be {\it approximately } valid. In this region,  it may be 
possible to match the solution of these low
temperature renormalization group equations to strong coupling calculations.
Effects not contained within the scheme of the perturbative
renormalization group, such as the instanton effects described by Belavin and
Polyakov\cite{BP} may be essential to reach a satisfactory description of this
model.

\section{Conclusions}

The results derived in this paper are perplexing. While
$(2+\epsilon)$-expansion has never been successful\cite{Cardy} in deriving the
critical properties of the Heisenberg model, it has been useful in providing a
conceptual framework. In contrast, the expansion in $(4-d)$ of the
$\phi^4$-field theory has been both quantitatively and conceptually useful. To
us, the results derived raise serious doubts about the  usefulness 
of the $(2+\epsilon)$-expansion, because it is virtually certain that the phase
transition in the Heisenberg model for $d>2$ is described by two relevant
operators, the temperature and the magnetic field and not by infinitely many
relevant operators. This conviction is supported by precise finite size scaling
analysis as well as theoretical work based on the expansion around $d=4$. 

It might be argued that because the high gradient
operators do not feed into the equation for the temperature, the disordering
transition is well-described by the conventional analysis\cite{Hikami}. This
argument has little force as an infinite number of parameters must be fine tuned
to be zero. On the other hand, for small $\epsilon$, the pathological effect of
the high gradient operators will be felt at very long length scales. Thus, the
conventional analysis may be approximately valid for shorter length scales. More
precise statements are difficult to make. It is also possible that
higher loop corrections may make the anomalous dimension of the high gradient
operators negative. While this cannot be ruled out, such a situation will still
imply rather unusual properties of the
$(2+\epsilon)$-expansion, if it is necessary to go to very high orders to
eliminate the pathological behavior of this expansion.

Based on our present understnding of the $O(N)$ model for which there are many
precise theoretical checks, we are forced to end on a negative note. In the
theory of Anderson localization, $(2+\epsilon)$-expansion has been used to derive
a number of interesting conclusions concerning the distribution of the
fluctuation of the moments of the conductance\cite{Altshuler}. In fact, it is
precisely the context in which the anomalous behavior of the high gradient
operators\cite{Kravstov} was first discovered. While these effects may 
exist, it is difficult to accept them on the basis of the
$(2+\epsilon)$-expansion. The most recent work of Dupr\'e\cite{Dup},  in which a
numerical simulation of a hyperbolic superplane model is carried out,  indicates
that there is only one relevant operator at the Anderson localization transition,
as was conceived originally by Abrahams {\em et al.\/}\cite{Abrahams}. 

In conclusion, we note that the $(2+\epsilon)$-expansion 
does not seem to reproduce what we believe to be the correct behavior of the
lattice Heisenberg model. It is possible that non-perturbative effects involving
topological excitations are important. Such considerations have been discussed
in the past\cite{Halperin,Cardy} and have been  emphasized recently
in careful numerical simulations\cite{Dasgupta}.

\section*{acknowledgement}
This work was supported by a grant from the National Science Foundation, grant: 
DMR-9531575 and by the Division of Material Sciences, U.S. Department of
Energy, under contract DE-AC02-76CH00016.
\appendix 
\section{}
In this appendix we provide an account of  the calculation of Eqs.
(\ref{rs1.1a}-\ref{rs1.2a}).
For convenience, we  define the dimension of the fields as follows
\begin{equation}
[\partial \xi]=[\xi]=0, \;\;\; [A_{\mu}]=[\partial\psi]=1 .
\end{equation}
Thus, any average of the field $\xi(x)$ will have dimension zero.

\begin{figure}[htb]
\centerline{\epsfxsize 9cm
\epsffile{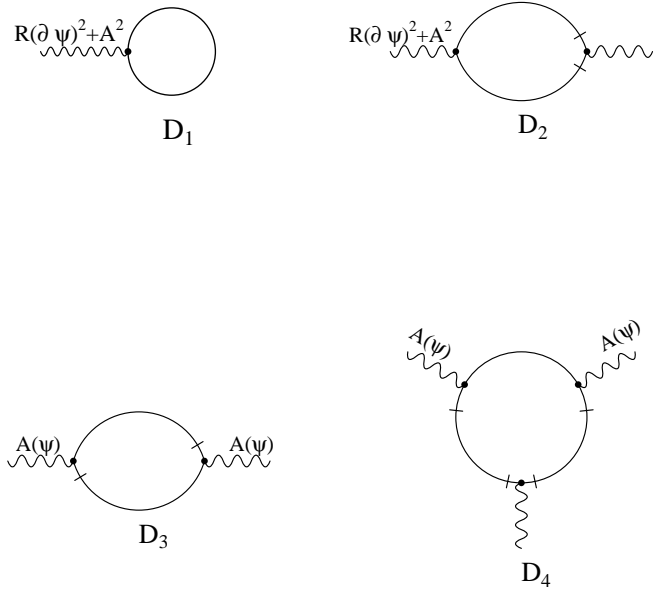}
}
\caption{One-loop diagrams contributing to the expectation
value
$\langle [G_{\mu\nu}]_{\xi^2} \rangle $.
The wiggly lines correspond to 
functions of the background field. Slashed  internal lines 
represent derivatives of the quantum field $\xi$.}
\label{fig:one-loop1}
\end{figure}

The one-loop divergences of the operator
$G_{\mu\nu}(\pi)$ are obtained from the normal
coordinate expansion through second order in the
field  $\xi(x)$
\cite{Alvar81}. Making use of Eqs. (\ref{Exp1}-\ref{Exp2})
the normal coordinate expansion for the action
and the operator are
\begin{eqnarray}
[S(\pi)]_{\xi^2}
 &=&
              \frac{1}{2}\int dx  \left[
              (  \partial_{\mu}\xi^{a}+ A_{\mu}^{ab}\xi^{b}  )
              (   \partial_{\mu}\xi^{a}+ A_{\mu}^{ab}\xi^{b}  ) +
        R_{mabn}\partial_{\mu}\psi^{m}\partial_{\mu}\psi^{n}\xi^{a}\xi^{b}
                           \right] \;\; ,   \\
\left[G_{\mu\nu}(\pi) \right ]_{\xi^2}
 &  =  &
   \left[ (  \partial_{\mu}\xi^{a}+ A_{\mu}^{ac}\xi^{c}  )
(   \partial_{\nu}\xi^{a}+ A_{\nu}^{ab}\xi^{b}  ) +
 R_{mabn}\partial_{\mu}\psi^{m}\partial_{\nu}\psi^{n}\xi^{a}\xi^{b}\right]
     \;\; .
\end{eqnarray}
Since the operator $G_{\mu\nu}(\pi)$ has dimension $d_{G}=2$,
it can only lead to divergent terms of the same dimension.
Then, by  inspection, we find that the one-loop divergence of
the operator $G_{\mu\nu}(\pi)$ is given by
\begin{equation}
\langle \left[G_{\mu\nu}(\pi) \right ]_{\xi^2} \rangle_{1L} =
D_1+D_2+D_3+D_4 \; \; ,
\end{equation}
where the divergent
contractions are:
\begin{eqnarray}
D_1 &=&
[R_{ma_{1}a_{2}n}\partial_{\mu}\psi^{m}\partial_{\nu}\psi^{n} +
A_{\mu}^{ca_1}A_{\nu}^{ca_2}]
\langle \xi^{a_{1}}(x)\xi^{a_{2}}(x) \rangle  \;\; ,    \\
D_2 &=& -\frac{1}{2} [
R_{ma_{1}a_{2}n}\partial_{\alpha}\psi^{m}\partial_{\alpha}\psi^{n} +
A_{\alpha}^{ca_{1}}A_{\alpha}^{ca_{2}}]
\langle \xi^{a_{1}}(x)\xi^{a_{2}}(x)\partial_{\mu}\xi^c (y)
                    \partial_{\nu}\xi^c(y) \rangle] \;\; , \\
D_3 &=& -A^{ab}_{\mu} A^{dc}_{\alpha}
\langle \xi^c(x) \partial_{\alpha} \xi^d (x)
\xi^b(y) \partial_{\nu} \xi^a (y) \rangle + (\mu \leftrightarrow \nu)
\;\; ,\\
D_4 &=& \frac{1}{2}
A^{b_2 b_1}_{\gamma} A^{c_2 c_1}_{\beta}
\langle
\xi^{b_1}(x) \partial_{\gamma} \xi^{b_2} (x)
\xi^{c_1}(y) \partial_{\beta} \xi^{c_2} (y)
\partial_{\mu}\xi^d (z) \partial_{\nu}\xi^d(z) \rangle \;\; .
\end{eqnarray}
The diagrams associated with these contractions
 are shown in Fig. ~\ref{fig:one-loop1}.

Note that  the divergences of these diagrams  are logarithmic,
so we can safely set the external momenta to zero. Direct and
straighforward calculations give the following results:
\begin{eqnarray}
D_1 &=&I [
R_{mn}\partial_{\mu}\psi^{m}\partial_{\nu}\psi^{n} +
A_{\mu}^{ca}A_{\nu}^{ca}     ]   \;\; ,\\
D_2 &=& - \frac{1}{2} I \delta_{\mu\nu} [
R_{mn}\partial_{\alpha}\psi^{m}\partial_{\alpha}\psi^{n} +
A_{\alpha}^{ca}A_{\alpha}^{ca}     ]  \;\; , \\
D_3 &=& -2I A^{ab}_{\mu} A^{ab}_{\nu}   \;\; ,\\
D_4 &=&\frac{1}{2} I [2 A^{ab}_{\mu} A^{ab}_{\nu} +
    \delta_{\mu\nu} A^{ab}_{\alpha} A^{ab}_{\alpha} ] \;\; ,
\end{eqnarray}
where $I=-\frac{1}{2\pi \epsilon}$.
Thus,
\begin{equation}
\langle
[G_{\mu\nu}]_{\xi^2} \rangle_{1L} =
[R_{mn}\partial_{\mu}\psi^{m}\partial_{\nu}\psi^{n} -
   \frac{1}{2} \delta_{\mu\nu}
R_{mn}\partial_{\alpha}\psi^{m}\partial^{\alpha}\psi^{n} ] \; .
\end{equation}
It should be noted that the terms that depend on $A_{\mu}^{ab}$ cancell out.
This is a consequence that the only possible covariant term,
$F_{\mu\nu}^{ab} e_{am} e_{bn}=
   R_{mnij} \partial_{\mu} \psi^{i} \partial_{\nu} \psi^{j}$,
is antisymmetric under the interchange 
between $\mu$ and $\nu$, and then, it can not appear in the right
hand side of the foregoing equation.
% is of the
%form $F_{\mu\nu}^{ab}F_{\mu\nu}^{ab}$, which has dimension
%$d_{F}=4$.

\begin{figure}[htb]
\centerline{\epsfxsize 9cm
\epsffile{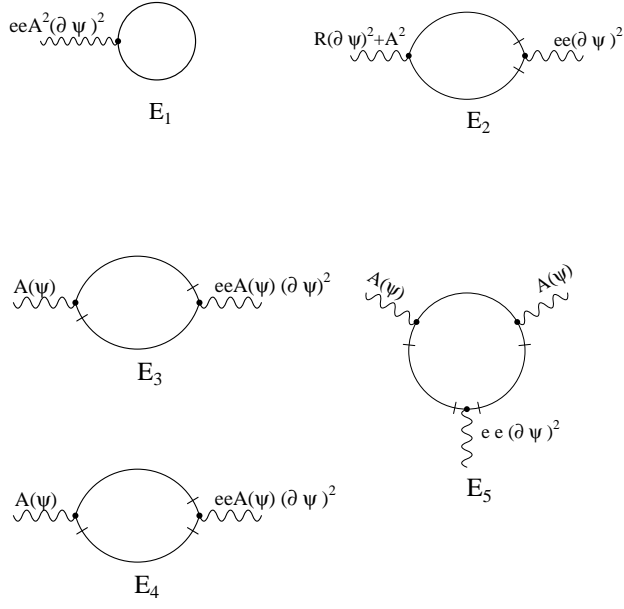}
}
\caption{ One-loop diagrams contributing to the expectation value
$\langle [G_{\mu\nu}]_{\xi^1} [G_{\alpha\beta}]_{\xi^1} \rangle$. }
\label{fig:one-loop2}
\end{figure}

Turning next  to the derivation of Eq. (\ref{rs1.2a}), we use
 Eq. (\ref{Exp1}) to obtain
\begin{eqnarray}
\left[ G_{\mu\nu}(\pi) \right]_{\xi^1}
[G_{\alpha\beta}(\pi)]_{\xi^1} &=&
e_{tb}e_{na}\partial_{\lambda}\psi^{t}\partial_{\nu}\psi^{n}
                  D_{\mu}\xi^{a}D_{\kappa}\xi^{b}+
 (\kappa \leftrightarrow \lambda,\mu \leftrightarrow \nu) + \nonumber\\
& &(\kappa \leftrightarrow \lambda)+
 (\mu \leftrightarrow \nu) \; ,  \label{app11}
\end{eqnarray}
where $D_{\mu}\xi^{a}$ is given in Eq. (\ref{Covar2}).
Using the fact that  the dimension of the operator
$G_{\mu\nu}G_{\kappa\lambda}$ is $d_{[O]}=4$, we can find that
the average of the {\it first} term of Eq. (\ref{app11}) is
given by
\begin{eqnarray}
\langle
e_{tb}e_{na}\partial_{\lambda}\psi^{t}\partial_{\nu}\psi^{n}
                  D_{\mu}\xi^{a}D_{\kappa}\xi^{b}
\rangle_{1L}
&=&
E_1+E_2+E_3+E_4+E_5   \;\; ,
\end{eqnarray}
where
\begin{eqnarray}
E_1 &=& e_{tb}e_{na}  \partial_{\lambda} \psi^{t} \partial_{\nu}\psi^{n}
A_{\mu}^{ad} A_{\kappa}^{be}
\langle \xi^d \xi^e \rangle \;\; ,  \\
E_2 &=&  -\frac{1}{2} e_{tb}e_{na}
 \partial_{\lambda} \psi^{t} \partial_{\nu}\psi^{n}
\left[
R_{pa_{1}a_{2}q}\partial_{\alpha}\psi^{p}\partial_{\alpha}\psi^{q} +
A_{\alpha}^{ca_{1}}A_{\alpha}^{ca_{2}}
\right]
\langle \xi^{a_1}(x) \xi^{a_2} (x)
 \partial_{\mu}\xi^a (y) \partial_{\kappa}\xi^b (y)
 \rangle \;\;, \\
E_3 &=& -e_{tb}e_{na} \partial_{\lambda} \psi^{t} \partial_{\nu}\psi^{n}
A_{\mu}^{ad} A_{\alpha}^{cf}
\langle \xi^f(x) \partial_{\alpha} \xi^c (x)
 \xi^d(y) \partial_{\kappa} \xi^b  (y)\rangle + \;\;
 (\mu \leftrightarrow \kappa, a \leftrightarrow b )
\;\; ,    \\
E_4 &=& - e_{tb}e_{na}
\partial_{\lambda} \psi^{t} \partial_{\nu}\psi^{n}
A_{\gamma}^{b_2 b_1}
\langle \xi^d(x) \partial_{\gamma} \xi^b   (x)
        \partial_{\mu}\xi^a i(y) \partial_{\kappa}\xi^b (y) \rangle \;\;,
           \\
E_5 &=& \frac{1}{2}
e_{tc_2}e_{nc_1} \partial_{\lambda} \psi^{t} \partial_{\nu}\psi^{n}
A_{\rho}^{b_2 b_1}  A_{\gamma}^{a_2 a_1}
\langle  \xi^{a_1} (x) \partial_{\gamma} \xi^{a_2} (x)
         \xi^{b_1} (y) \partial_{\rho} \xi^{b_2}  (y)
\partial_{\mu}\xi^{c_1} (z) \partial_{\kappa}\xi^{c_2} (z)
\rangle    \;\; .
\end{eqnarray}
The diagrams corresponding to these contraction are shown in
 Fig. ~\ref{fig:one-loop2}.
With the exception of $E_4$, all the contractions have dimension $4$
and are logarithmically  divergent,
so we can set the external momenta
to zero. In the case of $E_4$, the divergence is
linear and, thus,
proportional to the external momenta. In the Fourier space, the
calculation of $E_4$ yields
\begin{eqnarray}
\langle \xi^{b_1}\partial_{\gamma}\xi^{b_2}\partial_{\mu}\xi^a
                    \partial_{\kappa}\xi^c \rangle &=&
i  \delta^{b_1a}\delta^{b_2c}
\int dq \frac{(p+q)_{\mu} q_{\gamma}q_{\kappa}}
                  {q^2(q+p)^2} +
i    \delta^{b_1c}\delta^{b_2a}
\int dq \frac{(p+q)_{\kappa} q_{\gamma}q_{\mu}}
                  {q^2(q+p)^2}     \nonumber \\
&=&\frac{-i}{8\pi\epsilon} \{ \delta^{b_1a}\delta^{b_2c}
       \left[p_{\mu}\delta_{\kappa\gamma}-p_{\gamma}\delta_{\kappa\mu}
             -p_{\kappa}\delta_{\mu\gamma} \right]
 + \delta^{b_1c}\delta^{b_2a}
          \left[\mu \leftrightarrow \kappa \right] \} .
\end{eqnarray}
Direct calculations give the following results:
\begin{eqnarray}
E_1 &=& e_{tb}e_{na}
\partial_{\lambda} \psi^{t} \partial_{\nu}\psi^{n}
A_{\mu}^{ad} A_{\kappa}^{bd} I
\;\; , \nonumber   \\
E_2 &=&  -\frac{1}{2} \delta_{\mu\kappa}  I e_{tb}e_{na}
 \partial_{\lambda} \psi^{t} \partial_{\nu}\psi^{n}
[
R_{p\;\;q}^{\;ab\;}\partial_{\alpha}\psi^{p}\partial_{\alpha}\psi^{q} +
A_{\alpha}^{ca}A_{\alpha}^{cb}
 ]
  \;\; ,  \\
E_3 &=& 2 I e_{tb}e_{na}A_{\mu}^{ad} A_{\kappa}^{db}
\partial_{\lambda} \psi^{t} \partial_{\nu}\psi^{n}  \;\; ,
\nonumber  \\
E_4 &=& - \frac{1}{2} I e_{tb}e_{na}
\partial_{\lambda} \psi^{t} \partial_{\nu}\psi^{n}
(\partial_{\kappa}A_{\mu}^{ab}-\partial_{\mu}A_{\kappa}^{ab}) \;\;,
\nonumber  \\
E_5 &=& - \frac{1}{2}  I
e_{tc_2}e_{nc_1} \partial_{\lambda} \psi^{t} \partial_{\nu}\psi^{n}
[
A^{c_1 b}_{\alpha} A^{b c_2}_{\alpha} + A^{c_1 b}_{\mu} A^{b c_2}_{\kappa} +
A^{c_1 b}_{\kappa} A^{b c_2}_{\mu}
] \;\;.
\end{eqnarray}
Hence,
\begin{eqnarray}
\langle
e_{tb}e_{na}\partial_{\lambda}\psi^{t}\partial_{\nu}\psi^{n}
                  D_{\mu}\xi^{a}D_{\kappa}\xi^{b}
\rangle_{1L}
&=&
  -\frac{1}{2} \delta_{\mu\kappa}I
R_{ptnq}\partial_{\lambda} \psi^{t} \partial_{\nu}\psi^{n}
         \partial_{\alpha}\psi^{p}\partial_{\alpha}\psi^{q}
+ \nonumber \\
& & \frac{1}{2} I
e_{tb}e_{na} \partial_{\lambda} \psi^{t} \partial_{\nu}\psi^{n}
[(\partial_{\mu}A^{ab}_{\kappa} - \partial_{\kappa}A^{ab}_{\mu})
+ (A^{ac}_{\mu} A^{cb}_{\kappa} - A^{ac}_{\kappa}A^{cb}_{\mu})] \;\;.
\end{eqnarray}
Then, we find that the terms that depend on the gauge field $A_{\mu}^{ab}$
lead to
the tensor  $F_{\mu\nu}^{ab}$.
 In fact, it is not necessary to calculate
$E_5$ since from  gauge invariance and the result for $E_4$,
the coefficient that multiplies $F_{\mu\nu}^{ab}$ can be known.
Terms that are not
gauge invariant either cancel or are not divergent.
Making use of the definition of $F_{\mu\nu}^{ab}$ in
Eq. (\ref{identity}),
we obtain the  one-loop divergence
\begin{eqnarray}
\langle [ G_{\mu\nu}(\pi) ]_{\xi^1}
[G_{\kappa\lambda}(\pi)]_{\xi^1} \rangle &=&
\frac{1}{2} I  \partial_{\lambda} \psi^{t} \partial_{\nu}\psi^{n}
[- \delta_{\mu\kappa}
R_{ptnq}
         \partial_{\alpha}\psi^{p}\partial_{\alpha}\psi^{q} \;\; +
R_{tsmn}
         \partial_{\mu}\psi^{m}\partial_{\kappa}\psi^{n}
]  +
\nonumber \\
& & \;\;\; (\mu \leftrightarrow \nu) + (\kappa \leftrightarrow \lambda)+
    (\mu \leftrightarrow \nu ,\;\; \kappa \leftrightarrow \lambda) \; .
\end{eqnarray}

It may be noted that the results of this appendix are valid for arbitrary
Riemannian manifolds.

\section{}

In this appendix we discuss the calculation of the two-loop diagrams.
First we study the contribution of the diagrams of the type
(a) shown in Fig. ~\ref{twoloop1}.  
 Let $G$ denote the graph
in question, and  $g_{1}$ and $g_2$ the corresponding subgraphs.
Then, if $I_{G}$
denotes the Feynmann integral associated with $G$, we have that
\begin{equation}
I_{G} = I_{g_{1}}I_{g_{2}}, \nonumber
\end{equation}
Since $g_{i}$ are simply one-loop graphs, dimensional regularization tell us
that
\begin{equation}
I_{g_{1}}=\frac{a_{i}}{\epsilon} + b_{i},\nonumber
\end{equation}
where $a_{i}$ and $b_{i}$ are just constants that may depend on the external
momenta of the graphs $g_{i}$. Note that, if the
one-loop integral is finite, $a_{i}=0$.

If $T$ denotes the operation that selects the poles in $\epsilon$, the
application of the forest formula to G yields \cite{Coll84}
\begin{equation}
R(G)=I_{G}-T(I_{g_{1}})I_{g_{2}}  -T(I_{g_{2}})I_{g_{1}} \;\; ,
\end{equation}
where $R(G)$ denotes the overall divergence of the graph $G$.
When $I_{g_{1}}$ and $I_{g_{2}}$ are both divergent, the above formula
yields
\begin{eqnarray}
 R(G)&=&(\frac{a_{1}}{\epsilon} + b_{1})(\frac{a_{2}}{\epsilon} + b_{2})
                     -\frac{a_{1}}{\epsilon}(\frac{a_{2}}{\epsilon} + b_{2})
                     -\frac{a_{2}}{\epsilon}(\frac{a_{1}}{\epsilon} + b_{1})
                              \nonumber  \\
&=& -\frac{a_{1}a_{2}}{\epsilon^{2}} + {\rm finite} 
\end{eqnarray}
If  $I_{g_{2}}$ is finite, we obtain instead the result
\begin{eqnarray}
 R(G)&=&(\frac{a_{1}}{\epsilon} + b_{1})b_{2}
         -\frac{a_{1}}{\epsilon} b_{2} \nonumber \\
& = & {\rm finite}  
\end{eqnarray}
From these results, we conclude that the overall
 divergence of the two-loop graph in Fig ~\ref{twoloop1}(a) 
is proportional to
double poles in $\epsilon$.
 Thus, this type of graphs
do not lead to corrections to the anomalous dimension.

\begin{figure}[htb]
\centerline{\epsfxsize 9cm
\epsffile{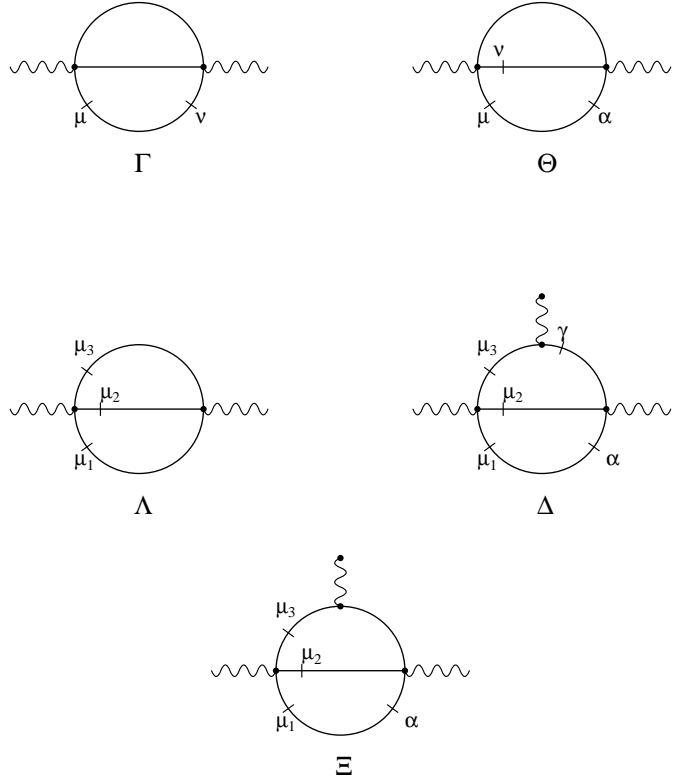}
}
\caption{Two-loop diagrams contributing to the expectation value 
$[G_{\mu_1\nu_1} G_{\mu_2\nu_2}]_{12}$ }
\label{appB.1}
\end{figure}

\begin{figure}[htb]
\centerline{\epsfxsize 9cm
\epsffile{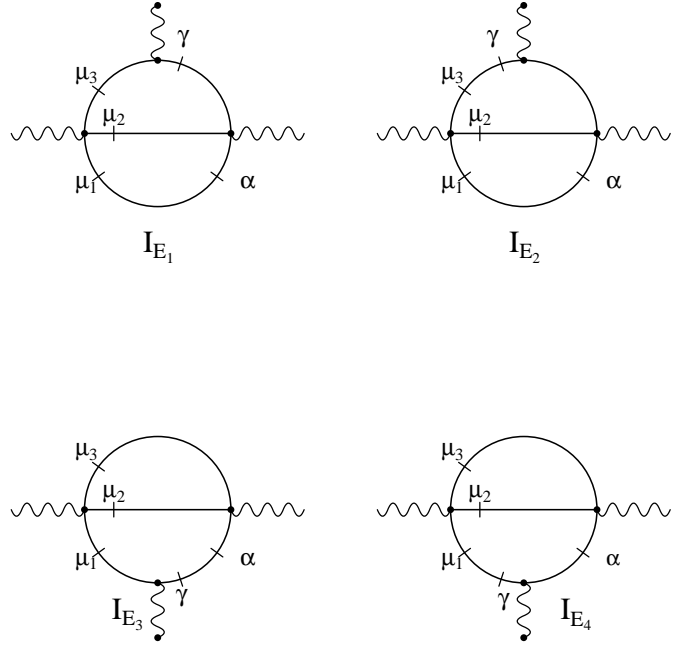}
}
\caption{The diagrams which arise from the calculation 
of the contraction $\Delta$}
\label{appB.2}
\end{figure}

Now we turn  to the calculation of the averages in Eqs.
(\ref{twoloopexp1}-\ref{twoloopexp2}).
 Since the calculation
of the averages is quite similar, we only show the calculation of  
the average 
\begin{equation}
[G_{\mu_1\nu_1}G_{\mu_2\nu_2}]_{12}=
\langle [G_{\mu_1\nu_1}]_{\xi^1}[G_{\mu_2\nu_2}]_{\xi^2}\rangle +
\langle [G_{\mu_2\nu_2}]_{\xi^1}[G_{\mu_1\nu_1}]_{\xi^2}\rangle 
\end{equation}
Making use of Eqs. (\ref{Exp1}-\ref{Exp2})
\begin{eqnarray}
[G_{\mu_1\nu_1}]_{\xi^1}[G_{\mu_2\nu_2}]_{\xi^2}
&=&
[e_{na} \partial_{\nu_1}\psi
D_{\mu_1}\xi^{a} + \mu_1 \leftrightarrow \nu_1 ] \times \nonumber \\
& & \;\;[e_{kb} e_{lb} D_{\mu_2} \xi^k D_{\nu_2} \xi^l +
R_{kcdl} \partial_{\mu_2} \psi^k \partial_{\nu_2} \psi^l
\xi^c \xi^d ]    
\end{eqnarray}
Substituing $D_{\mu}\xi^a$ (see Eq. (\ref{Covar2})), we obtain
\begin{equation}
[G_{\mu_1\nu_1}]_{\xi^1}[G_{\mu_2\nu_2}]_{\xi^2}=
\sum_{i=1}^{6} Q_{i}(\psi,\xi)  ,
\end{equation}
where
\begin{eqnarray}
Q_1(\psi,\xi)
 & \equiv &
 e_{n_{1}a}   \partial_{\nu_{1}} \psi^{n_1}
 \partial_{\mu_{1}} \xi^{a}
\partial_{\mu_2} \xi^{d} \partial_{\nu_2} \xi^{d} +
(\mu_1 \leftrightarrow \nu_{1} )  \label{Q1.exp}
 \;\; ,  \\
Q_2(\psi,\xi)
 & \equiv &
e_{n_{1}a} \partial_{\nu_{1}} \psi^{n_1}
  [
A_{\mu_1}^{ac_1} \xi^{c_1} \partial_{\mu_2} \xi^{d} \partial_{\nu_2} \xi^{d} +
A_{\mu_2}^{dc_2} \xi^{c_2} \partial_{\mu_1} \xi^{a} \partial_{\nu_2} \xi^{d} +
 \nonumber  \\
& &  \;\;\;
A_{\nu_2}^{dc_3} \xi^{c_3} \partial_{\mu_1} \xi^{a} \partial_{\mu_2} \xi^{d} +
(\mu_1 \leftrightarrow \nu_1)
   ]
      \;\; ,   \\
Q_3(\psi,\xi) & \equiv &e_{n_{1}a} \partial_{\nu_{1}} \psi^{n_1}
   [
A_{\mu_1}^{ac_1} A_{\mu_2}^{dc_2}  \xi^{c_1} \xi^{c_2} \partial_{\nu_2} \xi^{d}+
A_{\mu_1}^{ac_1} A_{\nu_2}^{dc_3}  \xi^{c_1} \xi^{c_3} \partial_{\mu_2} \xi^{d}+
\nonumber \\
& &     \;\;\;
A_{\mu_2}^{dc_2} A_{\nu_2}^{dc_3}  \xi^{c_2} \xi^{c_3} \partial_{\mu_1} \xi^{a}+
(\mu_1 \leftrightarrow \nu_1)
   ]
  \;\; , \\
Q_4(\psi,\xi) & \equiv &e_{n_{1}a} \partial_{\nu_{1}} \psi^{n_1}
A_{\mu_1}^{ac_1} A_{\mu_2}^{dc_2} A_{\nu_2}^{dc_3}
\xi^{c_1} \xi^{c_2} \xi^{c_3} +
(\mu_1 \leftrightarrow \nu_1)
   \;\; ,  \\
Q_5 (\psi,\xi) & \equiv &e_{n_{1}a} \partial_{\nu_{1}} \psi^{n_1}
 \partial_{\mu_2} \psi^{m_2} \partial_{\nu_2} \psi^{n_2} 
R_{m_2 b_1 b_2 n_2}
   \xi^{b_1} \xi^{b_2} \partial_{\mu_1} \xi^{a}  +     
(\mu_1 \leftrightarrow \nu_1)  \;\; ,  \\
Q_6 (\psi,\xi) & \equiv &e_{n_{1}a} \partial_{\nu_{1}} \psi^{n_1}
 \partial_{\mu_2} \psi^{m_2} \partial_{\nu_2} \psi^{n_2} 
R_{m_2 b_1 b_2 n_2}A_{\mu_1}^{ac_1}
\xi^{b_1} \xi^{b_2} \xi^{c_1}     + 
(\mu_1 \leftrightarrow \nu_1)\;\; . \label{Q6.exp}
\end{eqnarray}
In a similar fashion, the action $S$ through third order in
the field $\xi$ is given by
\begin{equation}
[S]_{\xi^2}+[S]_{\xi^3}= \int dx \left[S_1 +S_2 + S_3 + S_4 + S_5 \right]
\;\; ,
\end{equation}
where
\begin{eqnarray}
S_1(\psi,\xi) &\equiv& A_{\alpha}^{ab} \xi^{b} \partial^{\alpha} \xi^{a} \;\; ,
\label{S1.exp}   \\
S_2(\psi,\xi) &\equiv&
 \frac{1}{2} A_{\alpha}^{ab} A_{\alpha}^{ac} \xi^{b} \xi^{c}   \;\; , \\
S_3(\psi,\xi) &\equiv& \frac{1}{2} R_{i a_1 a_2 j} \partial_{\alpha} \psi^{i}
\partial^{\alpha} \psi^{j} 
 \xi^{a_1} \xi^{a_2}   \;\; ,\\
S_4(\psi,\xi) & \equiv& \frac{2}{3} R_{m a_1 a_2 a_3} 
\partial_{\alpha} \psi^{m} \xi^{a_1} \xi^{a_2}
 \partial\alpha \xi^{a_3}    \\
S_5(\psi,\xi)   & \equiv& \frac{2}{3} R_{m a_1 a_2 a_3} A_{\alpha}^{a_3 c}
 \partial_{\alpha} \psi^{m} \xi^{a_1} \xi^{a_2} \xi^{c}   \;\; .
\label{S5.exp}
\end{eqnarray}
In principle, one has to consider all the divergent
two-loop diagrams
that arise from the contraction between 
 {\it one} of
the vertices
$Q_n$ and the vertices $S_m$.
 However, the number
of diagrams can be reduced 
by using the dimension of the operator
$G_{\mu_1\nu_1}G_{\mu_2\nu_2}$, which is $4$,
and by calculating only those terms proportional to
the derivative of the gauge field
$\partial A$.
Also, let us recall that we are only interested in the operators
defined in  Eq. \ref{cyclic}.
 In this manner 
the only
terms necessary   
in the calculation of  
$[ G_{\mu_1\nu_1}G_{\mu_2\nu_2}]_{12}$  are 
\begin{eqnarray}
W_1(\psi)  &=&  (-1) \langle Q_5(\psi,\xi)   S_4(\psi,\xi) \rangle \;\; ,\\
W_2(\psi)  &=&  (-1) \langle Q_5(\psi,\xi)   S_4(\psi,\xi) \rangle \;\; ,\\
W_3(\psi)  &=&  (-1) \langle Q_5(\psi,\xi)   S_4(\psi,\xi) \rangle \;\; ,\\
W_4(\psi) &=& \langle Q_5(\psi,\xi)  S_4(\psi,\xi)
                  S_4(\psi,\xi)  \rangle \;\; ,\\
W_5(\psi) &=& \langle Q_5(\psi,\xi)  S_4(\psi,\xi)
                  S_4(\psi,\xi)  \rangle \;\; .\\
\end{eqnarray}
From Eqs. (\ref{Q1.exp}-\ref{S5.exp}), we see that the contractions  
of interest are
\begin{eqnarray}
\Gamma_{\mu\nu}^{b_1 b_2 b_3; a_1 a_2 a_3} (x,y) & \equiv &
\langle \xi^{b_1} \xi^{b_2} \partial_{\mu} \xi^{b_3} (x)
 \xi^{a_1} \xi^{a_2} \partial_{\nu} \xi^{a_3} (y) \rangle \; , \\
\Theta_{\mu\nu\alpha}^{b_1 b_2 b_3; a_1 a_2 a_3}(x,y) & \equiv &
\langle \xi^{b_1} \partial_{\mu} \xi^{b_2} \partial_{\nu} \xi^{b_3} (x)
 \xi^{a_1} \xi^{a_2} \partial_{\alpha} \xi^{a_3} (y) \rangle \; , \\
\Lambda_{\mu\nu\rho}^{b_1 b_2 b_3; a_1 a_2 a_3} (x,y)  & \equiv &
\langle \partial_{\mu} \xi^{b_1} \partial_{\nu}
 \xi^{b_2} \partial_{\rho} \xi^{b_3} (x)
 \xi^{a_1} \xi^{a_2} \xi^{a_3}(y) \rangle \; , \\
\Delta_{\mu_1\mu_2\mu_3\alpha\gamma}^{c_1c_2c_3;a_1a_2a_3;b_1b_2}(x,y,z)
 & \equiv &
    \langle \partial_{\mu_{1}}\xi^{c_{1}}
    \partial_{\mu_{2}}\xi^{c_{2}}
  \partial_{\mu_{3}}\xi^{c_{3}}(z)
\xi^{a_{1}} \xi^{a_{2}} \partial_{\alpha}\xi^{a_{3}}(x)
\xi^{b_{1}} \partial_{\gamma} \xi^{b_{2}} (y)\rangle \label{Delta1}  ,  \\
\Xi_{\mu_1\mu_2\mu_3\alpha}^{c_1c_2c_3;a_1a_2a_3;b_1b_2}(x,y,z)
 & \equiv &
    \langle \partial_{\mu_{1}}\xi^{c_{1}}
    \partial_{\mu_{2}}\xi^{c_{2}}
  \partial_{\mu_{3}}\xi^{c_{3}}(z)
\xi^{a_{1}} \xi^{a_{2}} \partial_{\alpha}\xi^{a_{3}}(x)
\xi^{b_{1}} \xi^{b_{2}} (y)\rangle \; .
\end{eqnarray}
The diagrams associated with these contractions are shown in 
Fig. ~\ref{appB.1}. 
To illustrate the method of calculation,
we now study the contraction $\Delta$.  In Fourier space,
we find
\begin{eqnarray}
\Delta_{\mu_1\mu_2\mu_3\alpha\gamma}^{c_1c_2c_3;a_1a_2a_3;b_1b_2}
(l,p) &=&
    \{ [I_{E_{1}}(\mu_{1},\mu_{2},\mu_{3})
    \delta^{c_{3}a_{3}} \delta^{c_{2}a_{2}} \delta^{c_{1}b_{1}}
    \delta^{b_{2}a_{1}} +
    I_{E_{2}}(\mu_{1},\mu_{2},\mu_{3}) \delta^{(b_{1} \leftrightarrow b_{2})}
]
  \nonumber  \\
&  &   + [I_{E_{1}}(\mu_{2},\mu_{1},\mu_{3})
     \delta^{c_{3}a_{3}} \delta{c_{2}b_{1}} \delta^{b_{2}a_{2}} 
     \delta^{c_{1}a_{1}} +
     I_{E_{2}}(\mu_{2},\mu_{1},\mu_{3}) \delta^{(b_{1} \leftrightarrow b_{2})
}]
  \nonumber   \\
& &  +  [I_{E_{3}}(\mu_{2},\mu_{1},\mu_{3})
     \delta^{c_{3}b_{1}} \delta^{b_{2}a_{3}} \delta^{c_{2}a_{2}}
       \delta^{c_{1}a_{1}} +
    I_{E_{4}}(\mu_{2},\mu_{1},\mu_{3})\delta^{(b_{1} \leftrightarrow b_{2})}]
 \nonumber   \\
& &   + [a_{2} \leftrightarrow a_{3}] +
        [a_{3} \rightarrow a_{2};a_{3} \rightarrow a_{1}] \}
+ \nonumber \\
& &   \{ a_{1} \leftrightarrow a_{2}  \},
 \end{eqnarray}
where $I_{E_{i}}$ are  just the Feynmann integrals of the diagrams of Fig.
~\ref{appB.2}. 
For instance the graph $I_{E_{1}}$ yields
\begin{eqnarray}
I_{E_{1}}&=& -i\int dq dk
\frac{q_{\alpha}q_{\mu_{3}}(k-q)_{\mu_{2}}(k-l-p)_{\mu_{1}}(k-l)_{\gamma}}
{[(k-l)^2+m^2][(k-l-p)^2+m^2][(k-q)^2+m^2][q^2+m^2]} \nonumber
\end{eqnarray}
where we have included the infrared cutoff in the propagators. 
By power counting, the integral
is linearly divergent (the superficial degree of divergence is $\delta=1$).
Hence, by the Weinberg theorem\cite{Coll84}, the overall
 divergence must be also linear, {\it i.e.},
the subtraction of divergences
cancels all the non-linear divergences. Thus,
we can expand
the propagators of the previous expression
in terms of the external momenta $l$ and $p$, and
 neglect all
the terms except for the linear term. This yields
\begin{eqnarray}
 iI^{'}_{E_{1}} &=&  (2p+4l)_{\delta}
\int dq dk
\frac{k_{\delta}k_{\mu_{1}}k_{\gamma}(k-q)_{\mu_{2}}q_{\mu_{3}}q_{\alpha}}
     {(k^2+m^2)^3(q^2+m^2)} \nonumber \\
 & & -l_{\gamma} \int dq dk
\frac{k_{\mu_{1}}(k-q)_{\mu_{2}}q_{\mu_{3}}q_{\alpha}}
     {(k^2+m^2)^2(q^2+m^2)} \nonumber \\
& & -(l+p)_{\mu_{1}} \int dq dk
\frac{k_{\gamma}(k-q)_{\mu_{2}}q_{\mu_{3}}q_{\alpha}}
     {(k^2+m^2)^2(q^2+m^2)}  \;\; .
\end{eqnarray}
Now
the calculation is reduced to computing the logarithmic divergences
of the integrals of the right hand side of the foregoing equation.
One of the  ways to
calculate the divergence  of this type integral
is to write the most general tensor form compatible
with the symmetries of the integral, and then to make
all the possible contractions to obtain relations
between the coefficients. In this way we obtain
\begin{eqnarray}
iI^{'}_{E_{1}} &=&
(2p+4l)_{\alpha}[ u_{1}\delta_{\mu_{1}\gamma} \delta_{\mu_{2}\mu_{3}} +
 u_{2}(\delta_{\mu_{1}\mu_{2}} \delta_{\gamma\mu_{3}} +
       \delta_{\mu_{2}\gamma} \delta_{\mu_{1}\mu_{3}} ) ]
          \nonumber \\
& & +(2p+4l)_{\mu_{3}}
    [ u_{1}\delta_{\mu_{1}\gamma} \delta_{\mu_{2}\alpha}+
    u_{2}(\delta_{\mu_{1}\mu_{2}} \delta_{\gamma\alpha}+
       \delta_{\mu_{2}\gamma} \delta_{\mu_{1}\alpha} ) ]
        \nonumber \\
& & +(2p+4l)_{\mu_{2}}
 [ u_{3}\delta_{\mu_{1}\gamma} \delta_{\mu_{3}\alpha}+
 u_{2}( \delta_{\mu_{1}\mu_{3}} \delta_{\gamma\alpha}+
       \delta_{\mu_{3}\gamma} \delta_{\mu_{1}\alpha} ) ]
        \nonumber \\
& & +(2p)_{\gamma}
 [ u_{3}\delta_{\mu_{1}\mu_{2}} \delta_{\mu_{3}\alpha}+
 u_{1}( \delta_{\mu_{2}\mu_{3}} \delta_{\mu_{1}\alpha}+
       \delta_{\mu_{1}\mu_{3}} \delta_{\mu_{2}\alpha} ) ]
        \nonumber \\
& & +(l)_{\gamma}
 [ u_{4}\delta_{\mu_{1}\mu_{2}} \delta_{\mu_{3}\alpha}+
 u_{5}( \delta_{\mu_{2}\mu_{3}} \delta_{\mu_{1}\alpha}+
       \delta_{\mu_{1}\mu_{3}} \delta_{\mu_{2}\alpha} ) ]
        \nonumber \\
& & +(p)_{\mu_{1}}
 [ u_{6}\delta_{\gamma\mu_{2}} \delta_{\mu_{3}\alpha}+
 u_{7}( \delta_{\mu_{2}\mu_{3}} \delta_{\gamma\alpha}+
       \delta_{\gamma\mu_{3}} \delta_{\mu_{2}\alpha} ) ]
        \nonumber \\
& & +(l)_{\mu_{1}}
 [ u_{4}\delta_{\gamma\mu_{2}} \delta_{\mu_{3}\alpha}+
 u_{5}( \delta_{\mu_{2}\mu_{3}} \delta_{\gamma\alpha}+
       \delta_{\gamma\mu_{3}} \delta_{\mu_{2}\alpha} ) ],
\end{eqnarray}
where the coefficients $u_{i}$  \vspace{.3cm} are
\begin{equation}
u_{1}=\frac{1}{\Omega}[-6+\epsilon/2], \;\;
u_{2}=- \epsilon/ \Omega , \;\;
u_{3}=\frac{1}{\Omega}[6-5\epsilon/2] ,
\end{equation}
\begin{equation}
u_{4}=8\epsilon/ \Omega, \;\;
u_{5}=4\epsilon/ \Omega, \;\;
u_{6}=\frac{1}{\Omega}[-12+13\epsilon],
 \end{equation}
\begin{equation}
u_{7}=\frac{1}{\Omega}[12-5\epsilon],
\end{equation}
and $\Omega=1536\pi^2 \epsilon^2$.

\begin{figure}[htb]
\centerline{\epsfxsize 9cm
\epsffile{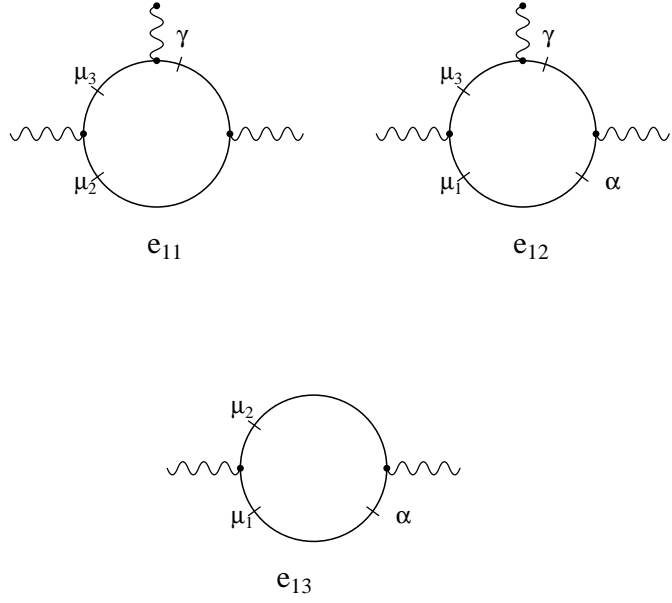}
}
\caption{The subgraphs associated with $I_{E_1}$.}
\label{appB.3}
\end{figure}

Let us now turn to the subtraction of subdivergences of the graph $I_{E_{1}}$.
 The
 three subgraphs of $I_{E_{1}}$ are shown in Fig. ~\ref{appB.3}.
Since the subgraph $e_{11}$ is finite, the application
of the Forest formula gives
\begin{equation}
R_{E_{1}}= I^{}_{E_{1}}-T(I_{e_{12}})I_{e_{13}}-T(I_{e_{13}})I_{e_{12}}
\end{equation}
After the subtractions are carried out, we find that
the overall divergence $R_{E_{1}}$ is identical to the integral
$I_{E_{1}}$ except for the sign of the $\frac{1}{\epsilon^2}$
term . 
The final result for 
$[G_{\mu_1\nu_1}G_{\mu_2\nu_2}]_{12}$ is then
\begin{eqnarray}
[ G_{\mu_{1}\nu_{1}}G_{\mu_{2}\nu_{2}} ]_{12}
& \equiv &
 \langle [G_{\mu_{1}\nu_{1}}]_{\xi^1}[G_{\mu_{2}\nu_{2}}]_{\xi^2} \rangle_{W}+
 \langle [G_{\mu_{1}\nu_{1}}]_{\xi^2}[G_{\mu_{2}\nu_{1}}]_{\xi^1} \rangle_{W}
        \nonumber  \\
 & = &
[E_{m_{1}n_{1},m_{2}n_{2}}
\Psi^{m_{1}n_{1}m_{2}n_{2}}_{\mu_{1}\,\nu_{1}\,\mu_{2}\,\nu_{2}}
+F_{pq,n_{1}n_{2}}
\delta_{\mu_{1}\mu_{2}}
\Psi^{n_{1}n_{2}}_{\nu_{1}\,\nu_{2}}\Psi^{p\,q}_{\alpha\alpha}]
        \nonumber \\
& &
+[\mu_{1} \leftrightarrow \nu_{1}]+
[\mu_{2} \leftrightarrow \nu_{2}]
+[\mu_{1} \leftrightarrow \nu_{1}, \mu_{2} \leftrightarrow \nu_{2}]+
        \nonumber \\
& &
G_{pq,mn}\Psi^{p\;q}_{\alpha\alpha}
[\delta_{\mu_{2}\nu_{2}}
\Psi^{m\;\;n}_{\mu_{1}\,\nu_{1}}+
\delta_{\mu_{1}\nu_{1}}
\Psi^{m\;\;n}_{\mu_{2}\,\nu_{2}}]   \;\; ,
\end{eqnarray}
where
\begin{eqnarray}
E_{m_{1}n_{1},m_{2}n_{2}}
 &=&
-\frac{\epsilon}{\Omega}
[48R_{n_{1}s}R^{s}_{\;\;\; n_{2}m_{1}m_{2}}+40R_{n_{1}(n_{2}p)q}
               R^{pq}_{\;\;\;\;\;m_{1}m_{2}}]
     \; , \label{res12.1} \\
F_{pq,n_{1}n_{2}} &=&
\frac{\epsilon}{\Omega}
[16R_{ps}R^{s}_{\;\;\; n_{1}n_{2}q}-4R_{p(n_{1}s)t}
               R^{st}_{\;\;\;\;n_{2}q}]+
[n_{1} \leftrightarrow n_{2}]
                    ,  \\
G_{pq,mn} &=&
-\frac{\epsilon}{\Omega}
[8R_{ps}R^{s}_{\;\;mnq}+16R_{p(ms)t}
               R^{st}_{\;\;\;\;nq}]+
[m \leftrightarrow n] \; \; . \label{res12.3} 
\end{eqnarray}
In Eqs. (\ref{res12.1}-\ref{res12.3}), we 
{\it only} show the single pole contribution,
and $(ab)$ denotes symmetrization in the indices
$a$ and $b$.
Note that this result is only valid for
  manifolds which satisfy  $\nabla_t R_{mnpq}=0$.

\section{}
In this appendix we write down our results.
To simplify the expressions, in this appendix we adopt the notation
\begin{equation} 
\Psi_{\mu_{1}\mu_{2} \cdots}^{m_{1}m_{2} \cdots} \equiv
\partial_{\mu_{1}}\psi^{m_{1}}\partial_{\mu_{2}}\psi^{m_{2}}\cdots.
\nonumber
\end{equation}
The expectation values of the terms of order $O(\xi^2)$ give the following
results:
\begin{eqnarray}
\langle [G_{\mu\nu}]_{\xi^2}\rangle
 & = &
-\frac{64\epsilon}{\Omega}R_{na_{1}a_{2}a_{3}}R_{m}^{\;\;\;(a_{1}a_{2})a_{3}}
 [ \Psi^{mn}_{\mu\nu} +  \Psi^{mn}_{\alpha\alpha} \delta_{\mu\nu}]
\; ,     \\
\langle [G_{\mu_{1}\nu_{1}}]_{\xi^1} [G_{\mu_{2}\nu_{2}}]_{\xi^1}\rangle
 & = &
[M_{m_{1}n_{1},m_{2}n_{2}}
\Psi^{m_{1}n_{1}m_{2}n_{2}}_{\mu_{1}\,\nu_{1}\,\mu_{2}\,\nu_{2}} +
N_{pq,n_{1}n_{2}}
\delta_{\mu_{1}\,\mu_{2}}
\Psi^{n_{1}n_{2}}_{\nu_{1}\,\nu_{2}}\Psi^{p\,q}_{\alpha\alpha}
]+
\nonumber  	 \\
& & 
[\mu_{1} \leftrightarrow \nu_{1}]+
[\mu_{2} \leftrightarrow \nu_{2}]
+[\mu_{1} \leftrightarrow \nu_{1}, \mu_{2} \leftrightarrow \nu_{2}]
        \;\;  ,
\end{eqnarray}
where
\begin{eqnarray}
M_{m_{1}n_{1},m_{2}n_{2}}    & =& 
-\frac{\epsilon}{\Omega}
[ 24R_{m_{1}(st)n_{1}}R_{m_{2}\;\;\;\;n_{1}}^{\;\;\;\;\;st\;\;\;\;} +
\frac{40}{3}R_{m_{1}(n_{1}s)t}
R_{m_{2}(n_{2}}^{\;\;\;\;\;\;\;\;\;\;\;s)t} 
 \nonumber	 \\
 &  &
\makebox[1cm] +\frac{32}{3}R_{m_{1}(n_{1}s)t}
R_{m_{2}(n_{2}}^{\;\;\;\;\;\;\;\;\;\;\;s)t}] +
[n_{1} \leftrightarrow n_{2}]  
   \;\; , \\
N_{pq,n_{1}n_{2}} &=& 
-\frac{\epsilon}{\Omega}
[ 12R_{m_{1}(st)n_{1}}R_{m_{2}\;\;\;\;n_{1}}^{\;\;\;\;\;st} +
\frac{4}{3}R_{m_{1}(n_{1}s)t}
R_{m_{2}(n_{2}}^{\;\;\;\;\;\;\;\;\;\;\;t)s} 
    \nonumber 	 \\
& &
\makebox[1cm] + \frac{32}{3}R_{m_{1}(n_{1}s)t}
R_{m_{2}(n_{2}}^{\;\;\;\;\;\;\;\;\;\;\;s)t}] +
[n_{1} \leftrightarrow n_{2}]	 \;\; .
\end{eqnarray}

Now we write the resuts for the  expectation values
of the terms of $O(\xi^3)$:
\begin{eqnarray}
\langle [G_{\mu\nu}]_{\xi^3}\rangle
 & = &
\frac{128\epsilon}{\Omega}R_{na_{1}a_{2}a_{3}}R_{m}^{\;\;\;\;(a_{1}a_{2})a_{3}}
 \Psi^{mn}_{\mu \, \nu}  \;\; ,  \\
\left[  G_{\mu_1\nu_1}G_{\mu_2\nu_2}  \right]_{12}
 & = &
[E_{m_{1}n_{1},m_{2}n_{2}}
\Psi^{m_{1}n_{1}m_{2}n_{2}}_{\mu_{1}\,\nu_{1}\,\mu_{2}\,\nu_{2}}
+F_{pq,n_{1}n_{2}}
\delta_{\mu_{1}\mu_{2}}
\Psi^{n_{1}n_{2}}_{\nu_{1}\,\nu_{2}}\Psi^{p\,q}_{\alpha\alpha}]
	\nonumber \\
& &
+[\mu_{1} \leftrightarrow \nu_{1}]+
[\mu_{2} \leftrightarrow \nu_{2}]
+[\mu_{1} \leftrightarrow \nu_{1}, \mu_{2} \leftrightarrow \nu_{2}]+
	\nonumber \\
& &
G_{pq,mn}\Psi^{p\;q}_{\alpha\alpha}
[\delta_{\mu_{2}\nu_{2}}
\Psi^{m\;\;n}_{\mu_{1}\,\nu_{1}}+
\delta_{\mu_{1}\nu_{1}}
\Psi^{m\;\;n}_{\mu_{2}\,\nu_{2}}]  \;\; ,
\end{eqnarray}
where
\begin{eqnarray}
E_{m_{1}n_{1},m_{2}n_{2}} 
 &=&    
-\frac{\epsilon}{\Omega} 
[48R_{n_{1}s}R^{s}_{\;\;\; n_{2}m_{1}m_{2}}+40R_{n_{1}(n_{2}p)q}
               R^{pq}_{\;\;\;\;\;m_{1}m_{2}}]
                 \;\; ,   \\
F_{pq,n_{1}n_{2}} &=& 
\frac{\epsilon}{\Omega}
[16R_{ps}R^{s}_{\;\;\; n_{1}n_{2}q}-4R_{p(n_{1}s)t}
               R^{st}_{\;\;\;\;n_{2}q}]+
[n_{1} \leftrightarrow n_{2}]
                   \;\; ,   \\
G_{pq,mn} &=& 
-\frac{\epsilon}{\Omega}
[8R_{ps}R^{s}_{\;\;mnq}+16R_{p(ms)t}
               R^{st}_{\;\;\;\;nq}]+
[m_{1} \leftrightarrow n_{1}] \;\;  .         
\end{eqnarray} 
Also,
\begin{eqnarray}
\langle [G_{\mu_{1}\nu_{1}}]_{\xi^1} [G_{\mu_{2}\nu_{2}}]_{\xi^1}
 [G_{\mu_{3}\nu_{3}}]_{\xi^1} \rangle & = &
\{ \Psi^{m_{1}m_{2}m_{3}n_{1}n_{2}n_{3}}_
{\mu_{1}\;\mu_{2}\;\mu_{3}\;\nu_{1}\;\nu_{2}\;\nu_{3}}
[ U_{m_{1}n_{1},m_{2}n_{2},m_{3}n_{3}}+
       \nonumber \\
& &
\;\;\;\;\;\;\; +(m_{1} \leftrightarrow m_{2};n_{1} \leftrightarrow n_{2}) + 
    (m_{1} \leftrightarrow m_{3};n_{1} \leftrightarrow n_{3})] +
        \nonumber  \\
& &
\;\;\;\;\;\;\; \Psi^{n_{1}n_{2}n_{3}}_{\nu_{1}\;\nu_{2}\;\nu_{3}} 
\Psi^{p\;q}_{\alpha\alpha} 
[\delta_{\mu_{1}\mu_{2}} \partial_{\mu_{3}} \psi^{m}
(V_{pq,mn_{1}n_{2}n_{3}})+           \nonumber   \\
& &
   \;\;\; +\delta_{\mu_{1}\mu_{3}} \partial_{\mu_{2}} \psi^{m}
(n_{2} \leftrightarrow n_{3})+
\delta_{\mu_{2}\mu_{3}} \partial_{\mu_{3}} \psi^{m}
(n_{1} \leftrightarrow n_{3})] \}+ \nonumber
 \\
&  & \;\;\;\;  +\{\mu_{i} \leftrightarrow \nu_{i}\} \;\; ,
  \end{eqnarray}
with
\begin{eqnarray}
 U_{m_{1}n_{1},m_{2}n_{2},m_{3}n_{3}}  &=&
\frac{4\epsilon}{\Omega}
\{ (R_{m_{1}(n_{1}t)n_{2}} + 
5R_{m_{1}(n_{2}t)n_{1}})R^{t}_{\;\;n_{3}m_{2}m_{3}} \nonumber \\
& &
+(n_{2} \leftrightarrow n_{3};m_{2} \leftrightarrow m_{3}) \},
        \\
V_{pq,mn_{1}n_{2}n_{3}}  &=&
\frac{2\epsilon}{\Omega}
\{2R_{p(n_{1}n_{2})s}R^{s}_{\;\;n_{3}qm}-3R_{m(n_{1}n_{2})s}
R_{p(n_{3}\;\;\;\;\;\;q}^{\;\;\;\;\;\;\;s)}+  \nonumber \\
& &
-4[R_{p(n_{1}n_{3})s}R^{s}_{\;\;n_{2}qm}+(n_{1} \leftrightarrow n_{2})]
     \nonumber \\
& &  \;\;\;\;-[R_{m(n_{3}s)n_{1}} + 5R_{m(n_{1}s)n_{3}})
R_{p(n_{2}\;\;\;\;\;\;q}^{\;\;\;\;\;\;\;s)}]
-[n_{1} \leftrightarrow n_{2}] \}   \; \; .
\end{eqnarray}

To find the results for the manifold $O(N)/O(N-1)$,
we substitute the curvature tensor, Eq. (\ref{curv2}),
in the previous results.
This yields,
\begin{eqnarray}
\langle [G_{\mu\nu}]_{\xi^2} \rangle
 &=&
-\frac{192 \epsilon}{\Omega}g_{mn}
[\delta_{\mu\nu}\Psi^{mn}_{\alpha\alpha} + \Psi^{mn}_{\mu\nu}],
      \\
\langle [G_{\mu_{1}\nu_{1}}]_{\xi^1} [G_{\mu_{2}\nu_{2}}]_{\xi^1}\rangle
&=&
-\frac{16\epsilon}{\Omega} P_{m_{1}n_{1},m_{2}n_{2}}
\Psi_{\mu_{1}\mu_{2}\nu_{1}\nu_{2}}^{m_{1}n_{1},m_{2}n_{2}}
         \nonumber  \\
 & & 
+\frac{12\epsilon}{\Omega}Q_{pqkl} \Psi^{p\;q}_{\alpha\alpha}
[(\delta_{\mu_{1}\mu_{2}} \Psi^{k\;l}_{\nu_{1}\nu_{2}} +
\delta_{\nu_{1}\mu_{2}} \Psi^{k\;l}_{\mu_{1}\nu_{2}}) +
(\mu_{2} \leftrightarrow \nu_{2})],
\end{eqnarray}
where
\begin{eqnarray}
P_{m_{1}n_{1},m_{2}n_{2}} & = &    
(18-4N)g_{m_{1}n_{1}}g_{m_{2}n_{2}}+
(9N-15)(g_{m_{1}m_{2}}g_{n_{1}n_{2}} +g_{m_{1}n_{2}}g_{n_{1}m_{2}}),
                   \\
Q_{pq,kl} &=& 
-10 g_{kl}g_{pq}+(11-3N)(g_{kp}g_{ql}+g_{kq}g_{pl}).   
\end{eqnarray}
Similarly, the $O(\xi^3)$ expectation values are 
\begin{eqnarray}
\langle [G_{\mu\nu}]_{\xi^3}\rangle &=& 
\frac{384\epsilon(N-2)}{\Omega} g_{mn}\Psi_{\mu\nu}^{mn},    \\
 \left[ G_{\mu_{1}\nu_{1}}G_{\mu_{2}\nu_{2}} \right]_{12}
&=&
\frac{48\epsilon}{\Omega} X_{m_{1}n_{1},m_{2}n_{2}} 
\Psi_{\mu_{1}\;\mu_{2}\;\nu_{1}\;\nu_{2}}^{m_{1}n_{1}m_{2}n_{2}}+
              \nonumber \\
& &
\frac{4\epsilon}{\Omega} Y_{pqkl} \Psi^{pq}_{\alpha\alpha}
[(\delta_{\mu_{1}\mu_{2}} \Psi^{kl}_{\nu_{1}\nu_{2}} +
\delta_{\nu_{1}\mu_{2}} \Psi^{kl}_{\mu_{1}\nu_{2}}) +
(\mu_{2} \leftrightarrow \nu_{2})]+
                      \nonumber \\
& &
-\frac{8\epsilon}{\Omega} Z_{pqkl}\Psi^{pq}_{\alpha\alpha}
[\delta_{\mu_{1}\nu_{1}}\Psi^{kl}_{\mu_{2}\nu_{2}}
   +\delta_{\mu_{2}\nu_{2}}\Psi^{kl}_{\mu_{1}\nu_{1}}];
%& & \nonumber
\end{eqnarray}
where
\begin{eqnarray}
 X_{m_{1}n_{1},m_{2}n_{2}}   & = &
(20N-66)g_{m_{1}n_{1}}g_{m_{2}n_{2}}+
(17-2N)(g_{m_{2}n_{1}}g_{n_{1}m_{1}}+g_{m_{1}m_{2}}g_{n_{1}n_{1}}),
               \\
Y_{pqkl}  
 & = & (32N-70)g_{kl}g_{pq}+ (23-10N)(g_{kp}g_{lq}+g_{pl}g_{kq}),
              \\
 Z_{pqkl} 
   & = & (26N-82)g_{kl}g_{pq}+ (23-4N)(g_{kp}g_{lq}+g_{pl}g_{kq}),
\end{eqnarray}
Finally,
\begin{eqnarray}
\langle [G_{\mu_{1}\nu_{1}}]_{\xi^1} [G_{\mu_{2}\nu_{2}}]_{\xi^1}
 [G_{\mu_{3}\nu_{3}}]_{\xi^1} \rangle & = &
\frac{4\epsilon}{\Omega} T_{m_{1}n_{1}m_{2}n_{2}m_{3}n_{3}}
\Psi^{m_{1}n_{1}m_{2}n_{2}m_{3}n_{3}}_
{\mu_{1}\;\nu_{1}\;\mu_{2}\;\nu_{2}\;\mu_{3}\;\nu_{3}}+
           \nonumber   \\
  & &           
\frac{2\epsilon}{\Omega}  S_{pqmnkl} \Psi^{p\;q}_{\alpha\alpha}
\times  \nonumber \\
 & &
\;\; \{ (\Psi^{m\;n}_{\mu_{3}\nu_{3}}
[\delta_{\mu_{1}\mu_{2}}\Psi^{k\;l}_{\nu_{1}\nu_{2}}  +
 \delta_{\nu_{1}\mu_{2}}\Psi^{k\;l}_{\mu_{1}\nu_{2}})  +
  (\mu_{2} \leftrightarrow \nu_{2})] +
           \nonumber   \\
  & &  
\;\; \Psi^{m\;n}_{\mu_{2}\nu_{2}}
[(\delta_{\mu_{1}\mu_{3}}\Psi^{k\;l}_{\nu_{1}\nu_{3}}  +
 \delta_{\nu_{1}\mu_{3}}\Psi^{k\;l}_{\mu_{1}\nu_{3}})  +
  (\mu_{3} \leftrightarrow \nu_{3})]+
           \nonumber   \\
  & &  
  \;\; \Psi^{m\;n}_{\mu_{1}\nu_{1}}
[(\delta_{\mu_{2}\mu_{3}}\Psi^{k\;l}_{\nu_{2}\nu_{3}}  +
 \delta_{\nu_{2}\mu_{3}}\Psi^{k\;l}_{\mu_{2}\nu_{3}})  +
  (\mu_{3} \leftrightarrow \nu_{3})]  \},
\end{eqnarray}
with
\begin{eqnarray}
T_{m_{1}n_{1}m_{2}n_{2}m_{3}n_{3}} &=&
-432g_{m_{1}n_{1}}g_{m_{2}n_{2}}g_{m_{3}n_{3}}
+144 \{
[g_{m_{3}n_{3}}(g_{m_{1}m_{2}}g_{n_{1}n_{2}} +g_{m_{1}n_{2}}g_{n_{1}m_{2}})]
                         \nonumber  \\
& &
\;\;\;\; +[m_{2} \leftrightarrow m_{3} ;n_{2} \leftrightarrow n_{3}]
+[m_{1} \leftrightarrow m_{3} ;n_{1} \leftrightarrow n_{3}] \}
                \nonumber  \\
& &
-54 \{
[g_{n_{1}n_{2}}(g_{m_{1}m_{3}}g_{m_{2}n_{3}} + g_{m_{2}m_{3}}g_{m_{1}n_{3}})]
  \nonumber  \\
& &
+[m_{2} \leftrightarrow m_{3} ;n_{2} \leftrightarrow n_{3}]
+[m_{1} \leftrightarrow m_{3} ;n_{1} \leftrightarrow n_{3}] \}
 \nonumber  \\
& &
-54 \{
g_{m_{3}n_{1}}g_{m_{2}n_{3}}g_{m_{1}n_{2}} +
g_{m_{1}n_{3}}g_{m_{2}n_{1}}g_{m_{3}n_{2}}  \}  \;\; ,
             \\
 S_{pqmnkl}    
       &=&
-104g_{mn}g_{pq}g_{kl}
+64g_{pq}(g_{mk}g_{nl} +  g_{nk}g_{ml})
 \nonumber  \\
& &
+48g_{mn}(g_{pk}g_{ql} +  g_{qk}g_{pl})
+16g_{kl}(g_{mp}g_{nq} +  g_{np}g_{mq})
 \nonumber  \\
& &
-19 \{
[g_{pn}(g_{mk}g_{ql}+g_{ml}g_{qk}) + g_{qn}(g_{mk}g_{pl}+g_{ml}g_{pk})]
+[m \leftrightarrow n]      \} \;\; .
\end{eqnarray}

\section{}
In this appendix we use conformal coordinates to simplify the 
two-loop results obtained in the previous appendix.
 Here, we  set  $\omega=24\pi^2 \epsilon$ and use the following notation:
\begin{eqnarray}
A({\pi}) &=& \partial_{+}\pi^{m} \partial_{+}\pi^{n}g_{mn}, \nonumber \\
B({\pi}) &=& \partial_{-}\pi^{m} \partial_{-}\pi^{n}g_{mn},  \nonumber \\
H({\pi}) &=& \partial_{+}\pi^{m} \partial_{-}\pi^{n}g_{mn}.  \nonumber
\end{eqnarray}
We denote the $O(\xi^n)$ term of the normal expansion of
$A({\pi})$,$B(\pi)$ and $H(\pi)$  by $a_{n}(\psi,\xi)$,
$b_{n}(\psi,\xi)$ and $h_{n}(\psi,\xi)$ respectively.
Straighforward manipulations yield
\begin{equation}
\langle a_2 \rangle_{2L} =-\frac{3\nu}{\omega}A({\psi})\;\;\;\; ,\;\;\;\;\;
\langle h_2 \rangle_{2L}=-\frac{9\nu}{\omega}H({\psi}), 
\end{equation}
\begin{equation}
\langle a_{1} a_{1} \rangle_{2L}=
-\frac{9\nu}{\omega}A({\psi})^2
\;\;\;\; ,  \;\;\;\;\;\;
\langle a_{1}h_{1} \rangle_{2L}=
-\frac{27\nu}{2\omega}A({\psi})H({\psi}),        
\end{equation}
\begin{eqnarray}
\langle a_1 b_1 \rangle_{2L}  &  =  &
-\frac{9}{\omega}[H^2({\psi})(\nu+1)+A({\psi})B({\psi})(\nu-1)]
     \;\; , \\
\langle h_1 h_1 \rangle_{2L}  &  =  &
-\frac{3}{2\omega}[H^2({\psi})(6\nu+2)+A({\psi})B({\psi})(3\nu-2)]
   \;\; .  
\end{eqnarray}
We now turn to the expectation values
 of the terms of order $O(\xi^3)$. We obtain
\begin{equation}
\langle a_3 \rangle_{2L}=
\frac{6\nu}{\omega}A({\psi}) \;\;\;\; , \;\;\;\;\;
\langle h_3 \rangle =\frac{6\nu}{\omega}H({\psi})  ,
\end{equation}
\begin{equation}
\langle [ a a  ]_{12}\rangle_{2L}=\frac{12\nu}{\omega}A({\psi})^2
\;\;\;\; , \;\;\;\;\;\;
\langle [a h]_{12} \rangle_{2L}=\frac{21\nu}{2\omega}A({\psi})H({\psi})
\; ,
\end{equation}
\begin{eqnarray}
\langle [a b]_{12} \rangle_{2L}  &  =  &
\frac{2}{\omega}[H^2({\psi})(4\nu+9)+A({\psi})B({\psi})(5\nu-9)],
       \\
\langle [h h]_{12} \rangle_{2L}  &  =  &
-\frac{3}{\omega}[H^2({\psi})(8\nu-3)+A({\psi})B({\psi})(3-2\nu)] \; ;
\end{eqnarray}
where $[ab]_{12} \equiv a_1 b_2 + a_2 b_1 $. Finally,
\begin{eqnarray}
\langle a_1 b_1 h_1 \rangle_{2L} &=&
\frac{14}{\omega}[H^3({\psi})- A({\psi})B({\psi})H({\psi})] \;\; ,  \\
\langle a_1 h_1  h_1 \rangle_{2L} &=&
\frac{2}{\omega}[H^2({\psi})A({\psi})- A({\psi})^2B_({\psi})] \;\; , \\
\langle a_1   a_1 b_1 \rangle_{2L} &=&
\frac{4}{\omega}[H^2({\psi})A({\psi})- A({\psi})^2B({\psi})] \;\; ,  \\
\langle h_1   h_1   h_1 \rangle_{2L} &=&
\frac{12}{\omega}[A({\psi}) B({\psi})H({\psi})-H^3({\psi})] \;\;  .
\end{eqnarray}
All other possibilities vanish.

\end{document}